\documentclass[sigconf]{acmart}
\definecolor{darkgreen}{rgb}{0,0.5,0}
\definecolor{orange}{rgb}{1,0.5,0}
\definecolor{teal}{rgb}{0,0.5,0.5}
\definecolor{darkpurple}{rgb}{0.5, 0, 0.5}
\definecolor{olive}{rgb}{0.6,0.6,0}

\newcommand{\rewrite}[1]{#1}

\newcommand {\systemname}{EchoLadder}

\AtBeginDocument{%
  \providecommand\BibTeX{{%
    \normalfont B\kern-0.5em{\scshape i\kern-0.25em b}\kern-0.8em\TeX}}}

\copyrightyear{2025}
\acmYear{2025}
\setcopyright{acmlicensed}\acmConference[UIST '25]{The 38th Annual ACM Symposium on User Interface Software and Technology}{September 28-October 1, 2025}{Busan, Republic of Korea}
\acmBooktitle{The 38th Annual ACM Symposium on User Interface Software and Technology (UIST '25), September 28-October 1, 2025, Busan, Republic of Korea}
\acmDOI{10.1145/3746059.3747659}
\acmISBN{979-8-4007-2037-6/2025/09}
\usepackage{xcolor} % 引入颜色包
\usepackage{soul} %高亮包
\usepackage{multirow} %表处理
\usepackage{graphicx}
\usepackage{subfigure}
\usepackage{bm}
\usepackage{fancybox}
\begin{document}

%\title{\systemname{}: A Large Vision Language Model Based Pipeline for AI-assisted Interactive Modification of Virtual Scenes}
%\title{\systemname{}: Immersive AI-Assisted Progressive Design of VR Scenes}
\title{\systemname{}: Progressive AI-Assisted Design of Immersive VR Scenes}

% \affiliation{%
\author{Zhuangze Hou}
\email{zhuanghou3-c@my.cityu.edu.hk}
\affiliation{%
  \institution{School of Creative Media, City University of Hong Kong}
  \city{Hong Kong}
  \country{China}}

\author{Jingze Tian}
\email{jingztian2-c@my.cityu.edu.hk}
\affiliation{%
  \institution{School of Creative Media, City University of Hong Kong}
  \city{Hong Kong}
  \country{China}}

\author{Nianlong Li}
\authornotemark[1]
\email{linianlong@iscas.ac.cn}
\affiliation{%
  \institution{Institute of software, Chinese Academy of Sciences}
  \city{Beijing}
  \country{China}}  

\author{Farong Ren}
\email{fr2305@nyu.edu}
\affiliation{%
  \institution{Steinhardt School of Culture, Education, and Human Development, New York University}
  \city{New York}
  \country{United State}}

\author{Can Liu}
\authornote{Co-corresponding authors}
\email{canliu@cityu.edu.hk}
\affiliation{%
  \institution{School of Creative Media, City University of Hong Kong}
  \city{Hong Kong}
  \country{China}}

% \affiliation{%
%   \postcode{43017-6221}}
%% You do not have to enter your paper ID

\renewcommand{\shortauthors}{Hou et al.}

\begin{abstract}
Mixed reality platforms allow users to create virtual environments, yet novice users struggle with both ideation and execution in spatial design. While existing AI models can automatically generate scenes based on user prompts, the lack of interactive control limits users' ability to iteratively steer the output.
In this paper, we present \systemname{}, a novel human-AI collaboration pipeline that leverages large vision-language model (LVLM) to support interactive scene modification in virtual reality. \systemname{} accepts users' verbal instructions at varied levels of abstraction and spatial specificity, generates concrete design suggestions throughout a progressive design process. The suggestions can be automatically applied, regenerated and retracted by users' toggle control. 
Our ablation study showed effectiveness of our pipeline components. 
Our user study found that, compared to baseline without showing suggestions, \systemname{} better supports user creativity in spatial design.
It also contributes insights on users' progressive design strategies under AI assistance, providing design implications for future systems. 

\end{abstract}





%%Teaser2.png

\begin{teaserfigure}
  \includegraphics[width=\textwidth]{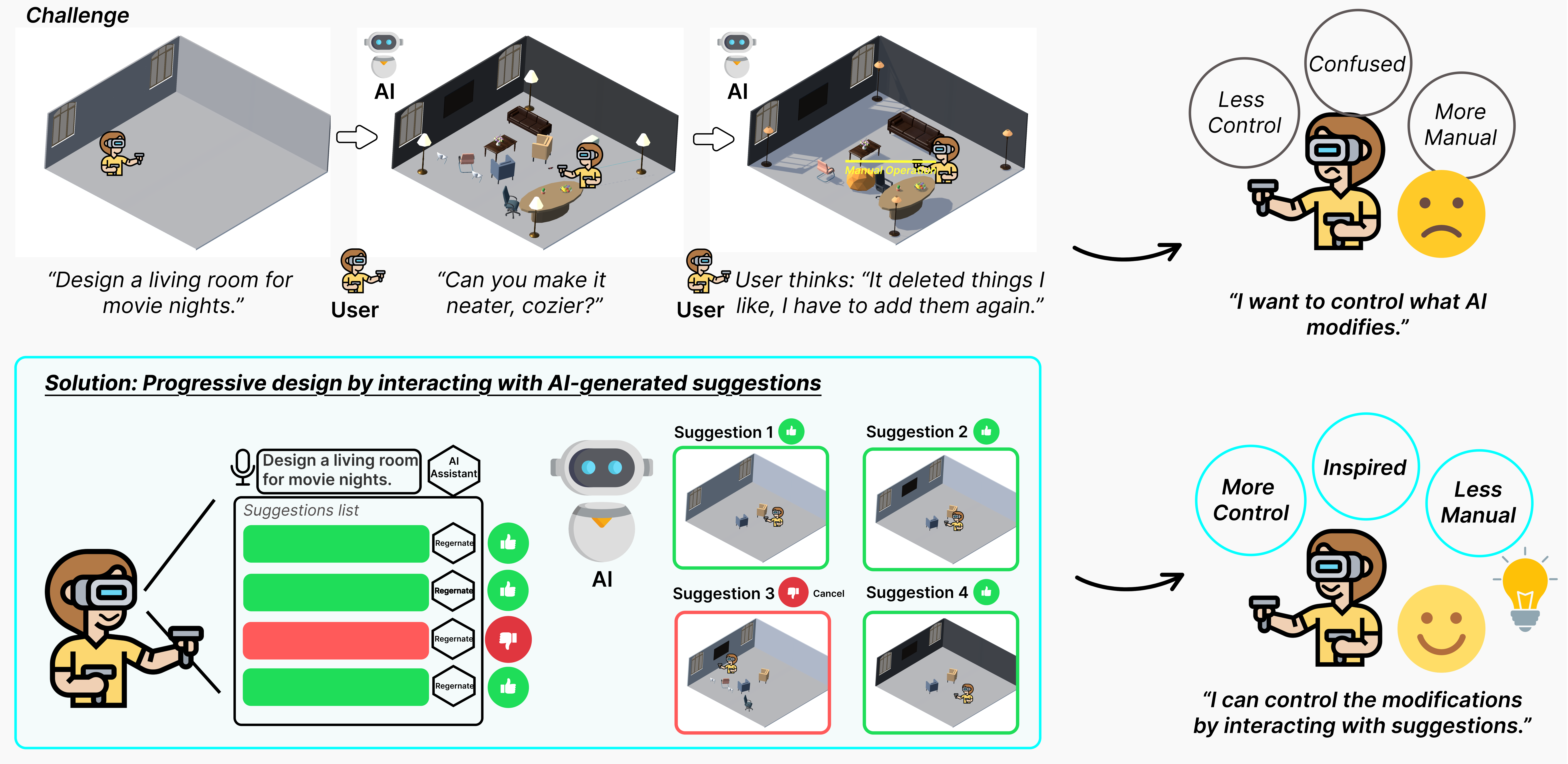}
  \caption{Immersive VR scene authoring with \systemname{}: \systemname{} makes the process of AI scene modification transparent by displaying interactable suggestion modules. Users can better control the AI modification process and modify the scene progressively. }
  \label{fig:teaser}
\end{teaserfigure}


\maketitle

\section{Introduction}

Recent advancements in generative 3D scenes, such as text-to-3D generation \cite{hong20233dllm, sun20233dgpt, yang2024holodeck} and LLM-based scene design \cite{Hou2024C2IdeasSC, zhang2024vrcopilot}, have introduced novel opportunities for AI-assisted VR authoring tools. By combining LLM understanding ability with VR authoring tools, these technologies enable users to craft intended immersive scenes more effectively. 
\rewrite{However, these automatic full-generation approaches predominantly follow a ``black-box'' generation model, limiting users to repeatedly re-generating or manually revising it post-hoc.} %Also, these methods may require excessive user effort, conflicting with a progressive authoring process needed for most design work.} 
%However, their integration also carries risks. The entire scene generation process is automatic, starting from a user’s descriptive prompt and ending with a generated scene due to the ``black-box'' nature of generative models.

Some recent works have integrated interactive methods to support users in iteratively building scenes. %to align user prompts with expectations. 
For instance, VRcopilot ~\cite{zhang2024vrcopilot} assists users in authoring VR layouts by allowing them to \rewrite{draw out areas or place wireframes to guide furniture generation,} thereby supporting a scaffolded process. %enhancing controllability. %Additionally, HOLODECK ~\cite{yang2024holodeck} facilitates 3D scene generation through multi-round human-AI conversations.
\rewrite{LLMR~\cite{de2024llmr} allows users to use natural language to modify objects in mixed reality scenes by integrating object parameters in the generation pipeline.} %in the scene and dynamically adapts them based on subsequent user instructions.}
%3D-GPT~\cite{sun20233dgpt}introduced a procedural pipeline for 3D modeling, which  integrates initial scene descriptions that are dynamically adapted based on subsequent instructions. 
Such solutions could fundamentally improve the support for iterative content creation. 
However, \rewrite{while much work is focused on improving AI pipelines for automatic modification, one important question remains: how to support user intervention in that process via effective interface solutions?} %most such prior works supporting content iteration focus on technical solutions of improving the AI pipeline. \zz{They do not explore two questions: ``how to support user intervention during AI-automated scene modification'' and ``how user interface helps user intervene automatic AI process''.}
%Little prior work on immersive scene generation focused on investigating and evaluating \zz{how user interface supports user intervene automatic AI process.} 

To fill this gap, we explore a design concept inspired by the effectiveness of ``chain of thought'' \cite{wei2022chain}, which is an established approach for improving the quality of AI generation. Previous works on AI-assisted writing identified benefits of exposing the thought process of generation to users. How would this translate to immersive spatial design? 
\rewrite{This research introduces a novel interface solution to support user intervention in AI-automated scene modification, to improve user agency and foster human-AI co-creation.}
%Building on these insights, %we extend the exploration of human-AI collaboration in VR authoring to the full task of immersive scene design.
% \zz{this research focuses on supporting AI-assisted scene modification by externalizing AI’s design reasoning and process through a modular user interface, in order to support user control and comprehension.}
%this research focuses on supporting AI-assisted scene modification, through the design concept of externalizing thought process into modular interactable UI, 

In this paper, we introduce a novel system named \systemname{}, based on a Large Vision-Language Model (LVLM), GPT-4o, to enable progressive and interactive scene construction in VR. \rewrite{Different from prior technical solutions~\cite{zhang2024vrcopilot, de2024llmr} executing users' concrete commands, \systemname{} interprets users' abstract instructions, combines scene images and object parameters to generate concrete modification suggestions. The suggestions are displayed as modular interactive widgets for users to selectively apply, undo, or regenerate. Once applied, the system executes a suggestion by modifying the scene. Users could view the direct visual effect and toggle to keep or retract the changes. This approach aims to provide both textual explanation and visual preview of each AI-generated step, while allowing selective and non-linear execution of them. } %\systemname{} automatically modifies and undoes scene modifications based on the user's execution of the suggestions, enabling a transparent and controllable immersive VR scene design experience.}
%\systemname{} interprets user instructions by combining scene images and object information, and generates modification suggestions by LVLM. Users can apply, undo, or regenerate these suggestions, enabling an automated, transparent, and controllable immersive VR scene design experience.

%\todelete{The system prototype was designed with three primary goals. First, to provide an automated scene modification pipeline for users without design experience, thereby lowering the entry barrier. Second, to decompose user instructions and provide design suggestions that inspire creativity and guide users throughout the design process. Third, to implement a progressive design interface offering users greater freedom and interaction in virtual scenarios design.}

We evaluated \systemname{} through two studies. An ablation study assessed the impact of removing each input component of the pipeline---\emph{visual input}, \emph{\rewrite{object parameters}}, and \emph{AI suggestions}---on the quality of generation results. The finding showed that the full input configuration used by \systemname{} achieved the best performance. \rewrite{A second user study compared \systemname{} with a baseline, which leverages the same pipeline for automatic scene modification but does not display intermediate suggestions to users. Our findings revealed our suggestion-based interface solution could better support user creativity and control, while leading to some distinct differences in design strategies.} 

% We make the following contributions:
% \zz{\begin{itemize}
%     \item We implement \systemname{}, a LVLM-based pipeline that integrates real-time scene understanding and vague user intents. Compared to other existing systems that directly generate outcomes, \systemname{} enables users to preview and interact with intermediate suggestions, supporting more user participation in the creative process. 
%     \item We propose a novel interaction model that foregrounds user intervention in AI-automated scene modification via intermediate, interactable suggestions, thereby enhance user control and creativity.
%     \item We evaluated the effectiveness of \systemname{} by ablation experiments and conducted a comparative user study against an end-to-end approach without intermediate AI-generated suggestions. Our empirical findings demonstrate the benefits of \systemname{} in supporting user creativity and diverse design strategies.
% \end{itemize}}
\rewrite{Our contributions are threefold:}
\begin{itemize}
   \item \rewrite{A novel interface solution supporting user intervention in iterative AI-automated authoring of VR scenes. It interprets users' natural language requests at any abstraction level and generates interactive suggestions for them to selectively apply. }% modification to support users iteratively building and modifying a VR environment by giving requests at any abstraction level. Compared to other existing systems that directly generate outcomes, our system allows users to preview and interact with intermediate suggestions, supporting more user participation in the creative process.
    \item \rewrite{An LVLM model-based AI pipeline integrating real-time scene understanding and semantic object retrieval to generate 3D scene modifications responding to users' natural language requests. The effectiveness of each pipeline component is validated by an ablation study. }
    %design of \systemname{} verified by an ablation study evaluating the quality of generation outcome.
    \item \rewrite{Empirical findings from a comparative user study evaluating \systemname{} against an AI-modification baseline, demonstrating the effectiveness of our suggestion-based design, and providing insights into how participants used the system. }%two different interaction styles based on our system: end-to-end generation versus suggestions participatory creation. 

\end{itemize}

\section{Related Work}

\subsection{AI-assisted scene generation}

There has been existing research on AI-assisted interior design, both in 2D and 3D.
Large models contribute a lot to 2D interior design.
For example, an Interactive Interior Design Recommendation System~\cite{Zhang_2023} based on reinforcement learning. The interaction with the user taps into the potential preferences of the homeowner and selects the appropriate initial design. 
Virtual Interior DESign system~\cite{le2023vides}, leveraging cutting-edge technology in generative AI to assist users in generating and editing indoor scene concepts quickly, given user text description and visual guidance. 
C2Ideas~\cite{Hou2024C2IdeasSC}, through large models to better automate the generation of interior color design schemes that are more consistent with users' ideas.

Before the bloom of LLMs, traditional AI has been well used for 3D scenes synthesis. 
For example, the traditional method of using CLIPGraphs~\cite{agrawal2023clipgraphs} can better estimate the position of objects in an indoor scene from a benchmark set of object categories. 
CompoNeRF~\cite{bai2023componerf}, which interprets complex text into editable 3D layouts and supports innovative multi-object composition. 
And a framework~\cite{9321177} for quickly synthesizing indoor scenes, measuring more reasonable relationships between objects through CSR, and generating various performations simultaneously in seconds. 
A system~\cite{10.1145/3130800.3130805} for adaptive synthesis of indoor scenes given an empty room and only a few object categories. It exploit a database of 2D floor plans to extract object relations and uses the similar plan references to create the layout of synthesized scenes. 

Benefiting from LLMs' strong ability in understanding and reasoning, some recent work on LLM-assisted 3D scene creation improves AI-assisted generation, making it more aligned with users' intentions. For example, Hong et al. proposed 3D-LLMs ~\cite{hong20233dllm} that take 3D point clouds and their features as input and perform a diverse set of 3D-related tasks. Additionally, Sun et al. developed 3D-GPT~\cite{sun20233dgpt}, which integrates three core agents (the task dispatch agent, the conceptualization agent, and the modeling agent) to conduct an instruction-draven 3D modeling and positions LLM as problem solvers. Moreover, Yang et al. developed a fully automatic system, HOLODECK~\cite{yang2024holodeck}, that can generate diverse 3D scenes with user customized styles.

While bringing benefits, however, these previous works mainly employ end-to-end generation approaches that automativally generate entire scene in one go, instead of allow users to intervene the generation process. Inspired by this, our work bridges the gap by allowing users to interactively iterate and modify the generated scenes, with the system creating suggestions and modifications based on the real-time status of the scenes.

\subsection{Immersive creativity support}
With the development of VR technology, the demand for creation of virtual scenes has become more and more abundant. A lot of previous work has focused on the immersive creativity support for VR scenes. 
The traditional Visual Worlds in Miniature metaphor~\cite{WIM} provides a user interface technique for creating 3D scenes from different perspectives.
Recently, there has also been a lot of work focused on improving the creative experience of VR immersion, Including creating more novel ways of interacting with VR/AR~\cite{ProtoAR, FlowMatic, MechARspace}, 
as well as exploring the topic of collaborative immersive creation~\cite{VRception,RemoteLab}. 
For example, VRGit~\cite{VRGit}, a new collaborative VCS that visualizes version history as a directed graph composed of 3D miniatures,  and enables users to easily navigate versions, create branches, as well as preview and reuse versions directly in VR. FlowMatic~\cite{FlowMatic}, an immersive authoring tool that raises the ceiling of expressiveness by allowing programmers to specify reactive behavior.
Different from existing creative support in VR scenes, we leverage Vision-LLMs to support the creativity and manipulation of content creation.
% The combination of virtual scene and real scene, and the immersive creation in multiple Spaces has become a topic that people are more and more willing to study

\subsection{AI-supported progressive scene crafting}

The \textit{progressiveness} of AI-supported content creation lies in its iterative refinement process, where users retain authority and guide the AI’s execution. This progressive approach has demonstrated significant benefits across multiple HCI studies, such as writing stories with AI suggesting plot, style, and tone to facilitate rapid creative iteration \cite{yuan2022wordcraft, 10.1145/3491101.3519873}. While the concept of \textit{progressiveness} in AI generation of 3D space has not been widely explored, recent works have made significant contributions by bridging LLMs and 3D content creation. For example, researchers have proposed pipelines to enhance LLMs' understanding of user design intent and generate style-aligned scenes, as seen in HOLODECK \cite{yang2024holodeck}. 3D-GPT \cite{sun20233dgpt} \rewrite{supports iterative natural language commands to generate and author a 3D scene, eg. changing the color of generated flowers}. Meanwhile, HCI research has focused on interactivity between users and LLMs to improve controllability. For instance, Zhang et al. introduced VRCopilot \cite{zhang2024vrcopilot}, an LLM-assisted authoring system that improves 3D layout generation by allowing users to scaffold or guide the LLM’s output via wireframing. Additionally, Torre et al. \cite{de2024llmr} presented the Large Language Model for Mixed Reality (LLMR), which supports scene understanding, task planning, self-debugging, and memory management, enabling users to create scene content iteratively by modifying object parameters. \rewrite{The novelty of EchoLadder lies in its focus on supporting user intervention in \emph{automated scene modification} while interpreting abstract user requests. Our AI pipeline integrates real-time full-scene understanding (vision + object-level parameters) and semantic-matching object retrieval to enable automatically adding objects to reasonable positions. While existing works directly execute one authoring process following user commands, EchoLadder uniquely introduces interactive suggestions as intermediate explanation, visualization and control support to bridge automated execution and manual adjustment for each authoring operation. } %\zz{However, they do not examine how to support user intervention in automated processes of AI. And, they execute specific user instructions for scene authoring, rather than understanding user intent based on general user instructions. Based on these gaps, we want to investigate user intervention in AI-supported progressive scene crafting and support abstract user instructions input.} 

\section{\systemname{}}

In this section, we present the system design of \systemname{}. \systemname{} is a novel AI-assisted VR scene design system that interprets user intent, generates context-aware design suggestions, and facilitates controllable, iterative construction in immersive environments. In \systemname{}, \textit{Labeling Module} leverages LVLM to automatically annotate 3D assets with accurate object-matching information, enabling faster and more precise object retrieval while eliminating the need for manual selection or browsing by the user. \textit{Generative Module} module understands natural language instructions, based on the current scene context (visual input and object parameters) to generate relevant suggestions for scene modification. It automatically handles object selection and editing, allowing users to freely express their ideas in language while the system performs spatial reasoning and executes the corresponding logic. Additionally, rather than executing user instructions directly, the system generates one or more suggestions for each instruction. Users can preview, apply, undo, or regenerate each suggestion individually, combining manual adjustments to iteratively construct the environment and refine their design decisions.

% \subsection{Scope}
We position \systemname{} within the domian of interior design, a representative and well-established application area for immersive authoring tools~\cite{zhang2023vrgit, zhang2024vrcopilot}, such as Home Design 3D VR\footnote{https://en.homedesign3d.net/vr} and IKEA Virtual Interior Designer\footnote{https://present.digital/ikea/}. However, \systemname{} is not limited to this domain and can be generalized to other spatial design tasks. The prototyping system developed in this paper contains 2156 3D models for interior design from the Unity Asset Store includes a wide range of furniture, decorations and textures. The scene modification operations supported by our system are derived from prior literature~\cite{zhang2024vrcopilot,VRGit} and existing VR application~\cite{homedesign3dvr}. These operations encompass common object-level tasks in interior design, including: adding 3D objects (\textit{Add}), modifying object positions (\textit{Move}), rotating (\textit{Rotate}), scaling (\textit{Scale}), changing colors (\textit{Color}), adjusting materials or styles (\textit{Material}), and removing objects (\textit{Destroy}).

\subsection{Interaction Design}
In this section, we describe interactions of \systemname{}.

\subsubsection{\textbf{Instruction Input}}

\begin{figure}
    \centering
    \includegraphics[width=1\linewidth]{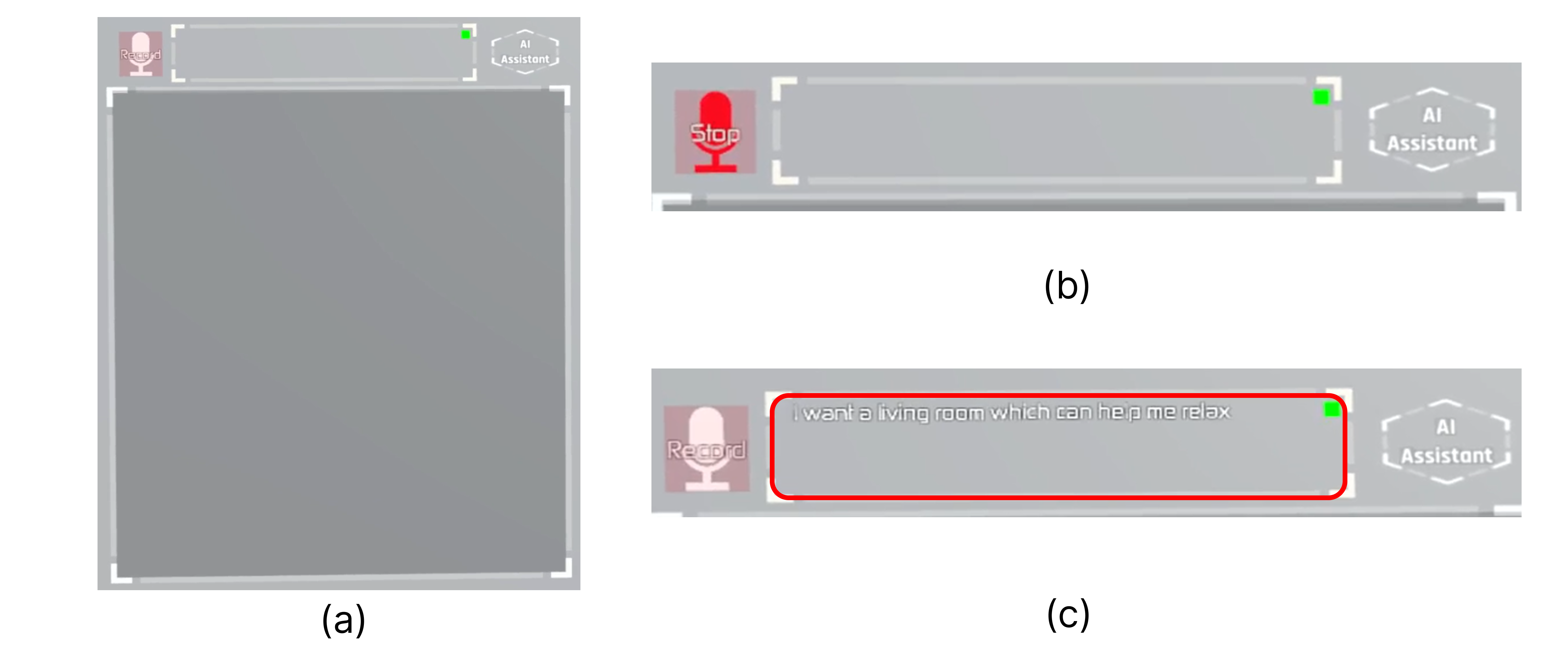}
    \caption{Voice input for user requests: (a) Initial state allows a user to input instruction after pressing the Mic button. (b) Mic button changes to Stop when accepting voice input. (c) Click Mic again to stop recording and check the transcribed user request. Clicking the AI Assistant button will start generating suggestions.}
    \label{VoiceInput}
\end{figure}

Considering the absence of physical keyboards in VR and the inefficiency of virtual typing~\cite{spiess2022direct, knierim2020opportunities}, we adopt intuitive voice~\cite{stedmon2011developing} for user instruction input. As shown in Figure~\ref{VoiceInput}, once the interaction interface is activated, users initiate voice recording by selecting the microphone icon using either the controller button or raycasting. The icon turns red to indicate active listening. After issuing a instruction, users select the icon again to stop recording, upon which the system transcribes the voice input into text and displays it on the interface. Based on the transcription accuracy, users can either proceed with the next operation or re-record the instruction.

\subsubsection{\textbf{User Interaction with Suggestions}}
\label{UserInteractionwithSuggestions}
\begin{figure}%[htbp]
\centering
	\subfigure[\label{Suggestions}] {\includegraphics[width=.21\textwidth]{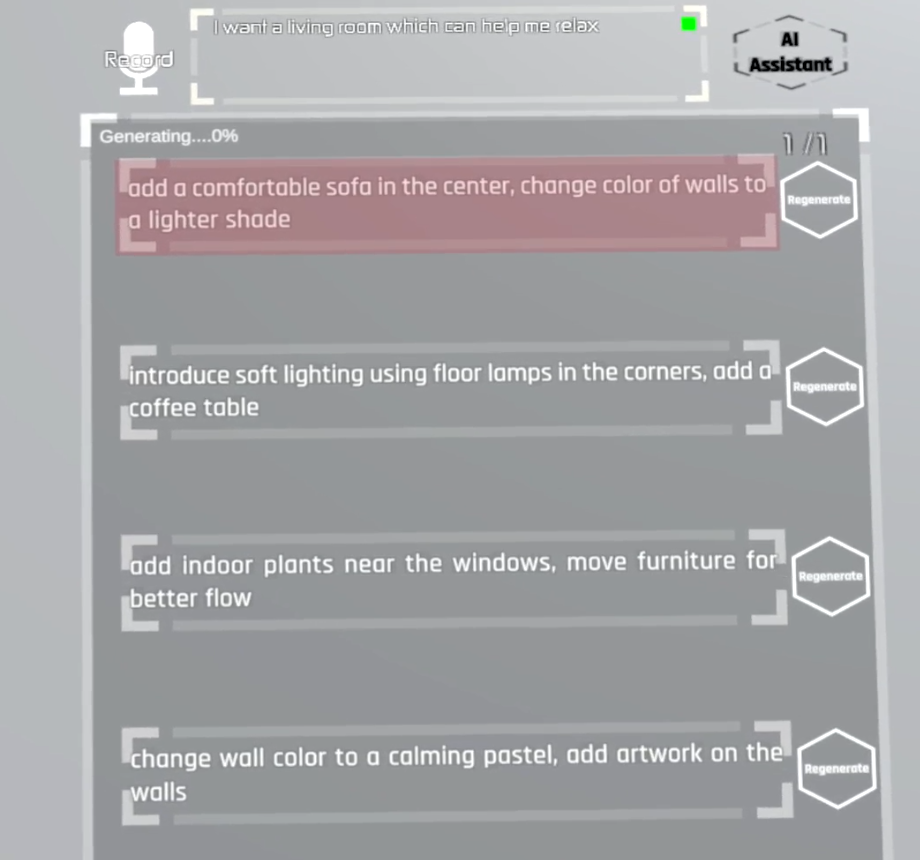}}
	\subfigure[\label{ThreeTypeSuggestions}] {\includegraphics[width=.23\textwidth]{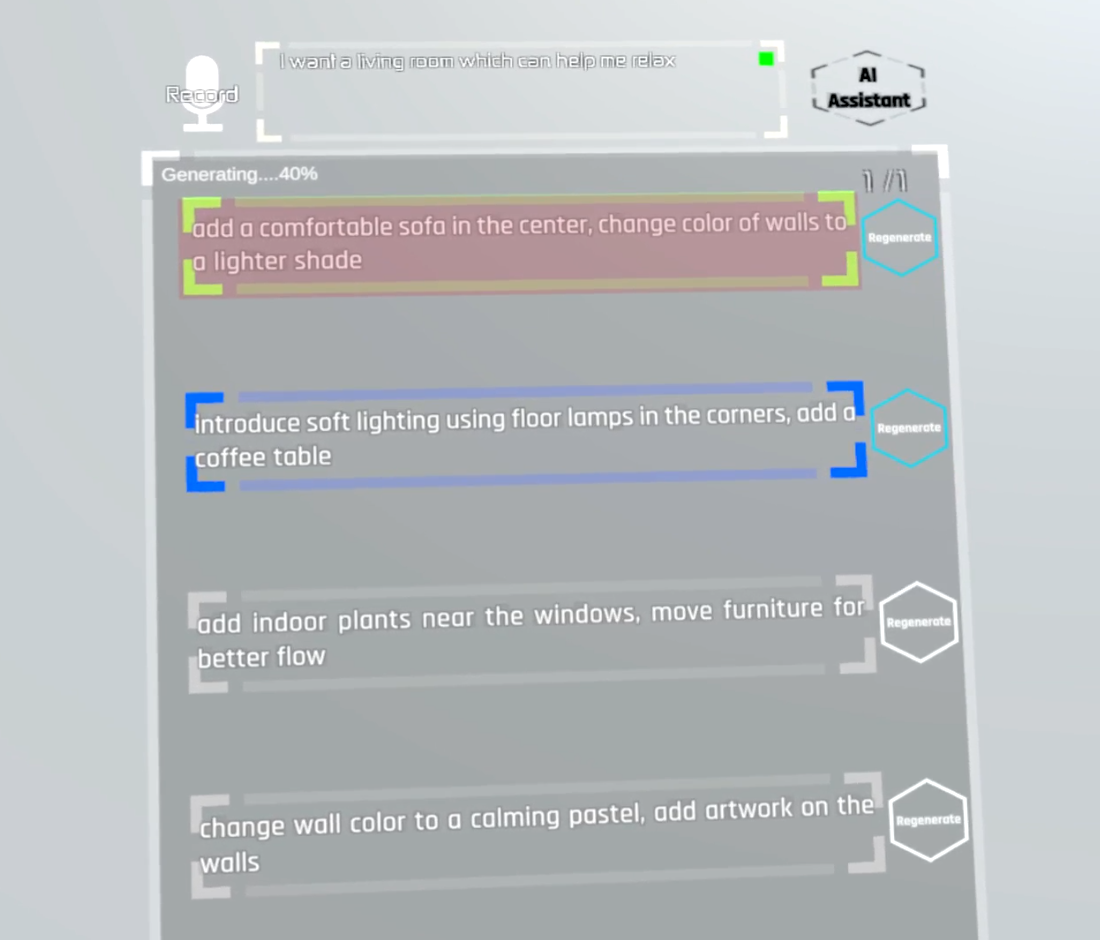}}
	\caption{AI-generated suggestions and their status in the interface.
(a) Aystem displays the generated suggestions after interpreting the user request. (b) Three statuses of the suggestions: White (Generation of spatial modification in progress), Blue (Generation completed, pending application), Green (Applied).}  
\label{InteractionwithSuggestions}
\end{figure}

 \rewrite{Our system uniquely introduces the display of interactive suggestions in order to show AI's reasoning process and selectively intervene in automated scene modification. }%Interaction with suggestions is an important component for our pipeline. This component allows users to view AI’s reasoning process, gain inspiration from it, and intervene in automated scene modifications, enabling users to control the modifications according to their own design intentions.}
 Once the user confirms the instruction, they can select ``AI Assistan'' button on the interface. The \textit{Generative Module} then parses the user instruction, object parameters and visual information to generate suggestions, which are individually listed on the interface (Figure~\ref{InteractionwithSuggestions}). Once suggestions are generated, the user is offered three interaction options.

\paragraph{\textbf{Browsing Suggestions.}}
% Users can read and understand the suggestions in order and decide whether to apply them. 
We provide two ways to help users browse suggestions, text reading and voice reading. Users can either read the textual content displayed in the interface or have the system read a suggestion aloud by clicking the joystick after scrolling down the list (by moving the joystick) to locate it.

\paragraph{\textbf{Apply and Undo Suggestion.}} 
Users can decide whether to apply a suggestion to preview its effect within the scene. Since the system requires time to translate suggestions into executable actions, we use colored borders to indicate the current processing status. As shown in Figure~\ref{ThreeTypeSuggestions}, a white border (\textit{Processing}) indicates that the AI is generating the corresponding actions in the background. During this phase, users can browse the suggestions but cannot apply them. Once the system completes the analysis and successfully generates the actions, the suggestion border turns blue (\textit{Pending Application}), indicating that it is ready to be applied to the scene. After the user confirms the application and the actions are successfully executed in the VR environment, the suggestion border turns green (\textit{Applied}). \rewrite{The user could view the post-modification effect in the VR environment. } %\zz{modifications of the suggestion to VR environment will actually be applied to the scene.}.
In addition, our system supports undoing applied suggestions. When users click on a previously applied suggestion, the system rolls back all modifications associated with that specific suggestion without affecting others. This immediate restoration of the VR scene enhances user control and flexibility.

\paragraph{\textbf{Regenerate Suggestion.}}
If users are not satisfied with the outcome of a generated suggestion, our system provides a suggestion regeneration function (via the ``Regenerate'' button next to each suggestion). This allows users to modify the result of a single suggestion without re-entering the original instruction. Upon triggering regeneration, the suggestion border turns white, and the \textit{Generative Module} regenerates the corresponding actions based on the current scene context.

\subsubsection{\textbf{Immersive Manual Authoring Interactions}}
The system provides manual operations to support users to customize the design more freely. The manual operations include object modification and object addition:

\paragraph{\textbf{Object Modification}}
The system provides the following manual operations for modifying virtual objects:

\begin{itemize}
\item Select objects using the ray on the right-hand controller of VR device.
\item Manipulate objects by moving them along the handle ray, rotating them, or adjusting their size.
\item Modify object properties, including color and material, or delete objects via the interface.
\end{itemize}

\begin{figure*}[h]
  \centering
   \includegraphics[width=1\linewidth]{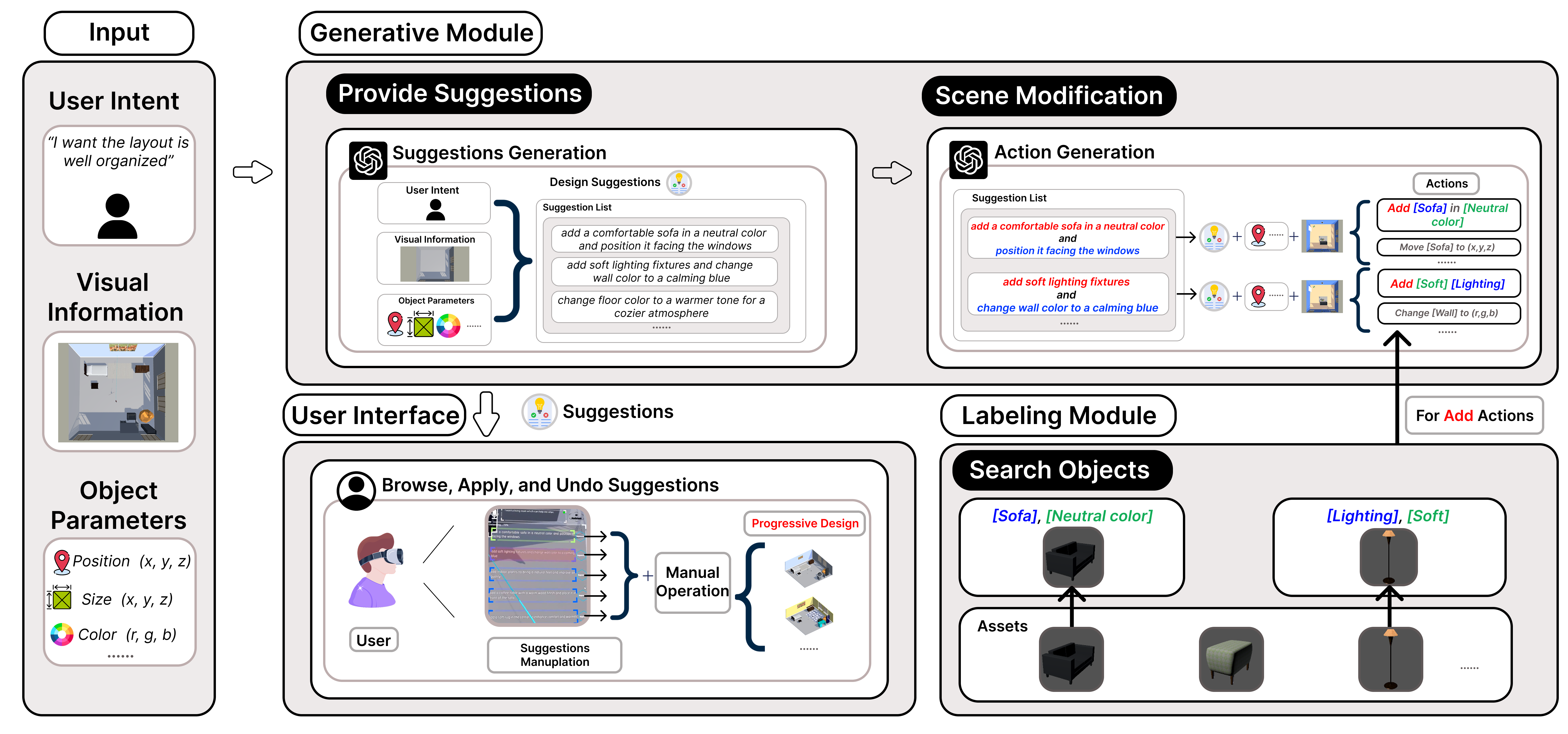}
  \caption{\systemname{}: In each iteration, the user provides a natural language instruction along with visual and object parameters as input. The \textit{Generative Module} then produces design suggestions and corresponding actions. These suggestions are presented in the interface, and \textit{Add} actions are linked to relevant assets by the \textit{Labeling Module}. Users can browses, apply, undo and regernerate suggestions, optionally making manual refinements. This process results in a progressively and iteratively updated scene.}
  \label{Pipeline}
\end{figure*}

\paragraph{\textbf{Manual Objects Addition.}}
The manual object addition menu enables users to browse and select object categories for searching 3D models. Selecting a category displays a list of available objects under that classification. Users can then add an object by clicking on its corresponding preview image.
It is important to note that the objects available for manual addition differ from those that can be automatically generated by the AI pipeline. This distinction ensures that users retain control over manual customization while leveraging AI-driven automation where necessary.

\subsection{Technical Architecture}

In this section, we introduce the architecture of \systemname{} in the immersive authoring system. There are two main modules: \textit{Labeling Module} and \textit{Generative Module}. 

\subsubsection{\textbf{Labeling Module}}
\rewrite{For the system to automatically add objects that match user's requests based on a mixture of considerations such as function and style, we designed a \textit{Labeling Module} to perform multi-dimensional, high-levelsemantic labeling for our 3D models. To the best of our knowledge, there were no publicly available datasets that meet our needs at the time we built the system.} %This module provides automatically 3D models annotations for system model assets and helps high-level semantic asset matching. 
To be specific, the \textit{Labeling Module} has two main parts: 

\paragraph{\textbf{Automatic Annotation of 3D Asset}}
Using LVLMs for image annotation has proven to be an effective approach~\cite{yang2024data, wu2025experimental}. This part includes four steps. First, thumbnails of all 3D models in system asset repository are extracted as the dataset for labeling. Second, the LVLM (GPT-4o) is used to iteratively generate open-vocabulary labels for each 3D model. These annotations include the model name, category, and a description summary (e.g., functions, colors), all inferred from the visual content of the thumbnails. Third, the generated annotations are structured in JSON format and stored as text files. \rewrite{The prompts and example of JSON file are available in Appendix~\ref{Details of Labeling Module}.} Finally, we manually reviewed the labeled content, with a particular focus on verifying the model description.

\paragraph{\textbf{Asset Matching.}} As shown in Figure~\ref{Pipeline}, when \textit{Generative Module} outputs an action such as ``\textit{add a comfortable sofa in a neutral color}'', it generates a corresponding description based on the user's intent and the specified action. It then passes the object name and description (e.g., ``\textit{sofa}'') to \textit{Labeling Module}. The \textit{Labeling Module} first identifies the appropriate object category using pre-assigned labels, and then employs a natural language processing algorithm (Sentence-BERT, or S-BERT) to retrieve the most relevant asset from the 3D model library---such as a light gray fabric sofa---that best matches the generated description.

\subsubsection{\textbf{Generative Module}}
In this system, the \textit{Generative Module} is responsible for interpreting user instructions, visual input and object parameters, subsequently generating scene modification suggestions along with specific execution actions. The entire process can be decomposed into the following four key steps:

\paragraph{\textbf{User Intent Recognition and Scene Comprehension.}} 
The \textit{Generative Module} first parses the input provided by the user (Appendix~\ref{Scene Understanding}). As shown in Figure~\ref{Pipeline}, the input consists of three primary components: user intent (\textit{Instruction}), visual information (\textit{Top view image of scene}), and object parameters (Parameters for each attribute of scene objects) 
%scene information (\textit{Object Parameters}), 
which includes objects' position, rotation, scale, color, and material (Style). 
%After giving these input to Generative AI user instruction along with vision information and scene information are transmitted to Generative AI (LVLM, GPT-4o).
After receiving these input, the \textit{Generative Module} analyzes the current scene by combining the visual and object parameters to identify objects, their attributes, and spatial relationships. Finally, through multimodal reasoning, the system determines whether the user’s instruction aligns with the existing scene and generates actionable suggestions accordingly. %\zz{In contrast to prior work~\cite{zhang2024vrcopilot,de2024llmr} that relied precise and specific user instructions, \textit{Generative Module} of \systemname{} interprets vague user instructions and intents based on visual information and object parameters for enhanced scene understanding.}
%user instruction along with vision information and scene information are transmitted to Generative AI (LVLM, GPT-4o). Second, AI combines vision information and scene information to analyze the current scene, identifying objects, attributes, and spatial relationships. Finally, through multimodal Reasoning, the AI determines whether the user’s instructions align with the existing scene and generates actionable suggestions accordingly.

\paragraph{\textbf{Generate Reasonable Scene Modification Suggestions.}} The \textit{Generative Module} formulates suggestions based on the previously analyzed information. As shown in Figure~\ref{Pipeline} (\textit{Suggestions Generation}), the module infers the most reasonable modifications by considering both the user's instruction and the current scene context---such as adjusting the position, size, color of objects, or adding new elements to the scene. These suggestions are structured in JSON format (Appendix~\ref{Suggestions Generation}) for downstream processing. The JSON-formatted suggestions are then parsed and converted to interactive buttons within the user interface. Finally, all suggested modifications are visually rendered, allowing users to browse and review them before making a decision.

%The AI enters the Generate Suggestions phase, first, it speculates the most reasonable modifications based on user instruction and scene, such as adjusting the position, size, color of objects or adding new elements to the scene. These suggestions are structured in JSON format for further processing. Second, the JSON-formatted suggestions are transmitted to the Suggestions Module within the Unity client interface, where they are parsed and presented for user review and selection. Finally, all suggested modifications are visually rendered in the Unity interface, allowing the user to confirm the changes or make additional refinements as needed.

% \begin{figure*}[h]
%   \centering
%    \includegraphics[width=0.85\linewidth]{Pic_table/AddSearch.png}
%   \caption{AI-assisted object addition: This diagram illustrates the process of selecting the most suitable object when adding an item to a VR scene. When an "Add" action is triggered in response to a user suggestion, AI generates a category (e.g., Furniture) and a detailed description of the desired object. The system then searches for matching objects within the Labeled Assets repository, filtering by category. Using S-Bert, the AI compares semantic similarity between the generated description and available objects, selecting the best-fit object.}
%   \label{AddSearch}
% \end{figure*}
\paragraph{\textbf{Translating Suggestions into Executable Actions.}} At this stage, the \textit{Generative Module} proceeds to the \textit{Action Generation} phase (Figure~\ref{Pipeline}, \textit{Action Generation}), where it transforms the generated suggestions into actions that can be directly applied to the VR scene. \rewrite{The system leverages LVLM to understand the scene, then parse each suggestion into action lists in JSON format (more details in Appendix~\ref{Actions Generation}).} Each action represents an executable command generated by the \textit{Generative Module} for a given suggestion, enabling automated scene modification. First, the module iterates through each suggestion, using the visual information, object parameters and suggestions as input, generates a set of concrete actions. These actions include \textit{Move}, \textit{Scale}, \textit{Rotate}, \textit{Color}, \textit{Style}, \textit{Delete}, and \textit{Add} operations. For example, a suggestion like ``\textit{add a neutral-colored sofa}'' could be translated into actions such as \textit{Add [sofa]} and \textit{Move [sofa] to $(x,y,z)$}. \rewrite{By interpreting both the object parameters and visual information, the system tries to place objects at appropriate positions in reasonable sizes.} For actions involving the addition of new objects, the \textit{Labeling Module} is invoked to retrieve a suitable object from the 3D model library and return it to the \textit{Generative Module}.
Finally, all generated actions are scructured in JSON format and associated with their corresponding suggestion. When the user applies a suggestion, all linked actions are executed in sequence with in the VR scene. 

\subsection{Implementation}
The prototype system presented in this paper was developed using Unity (version 2021.3.8f1c1) and integrated with SteamVR 2.8.0, enabling compatibility with both Meta Quest and HTC Vive headsets. The application featured advanced speech recognition and response capabilities through the integration of Whisper (Audio Model) and GPT-4o (Large Vision-Language Model, LVLM), allowing for natural and efficient user interaction. The system was deployed on a Windows 11 desktop equipped with an Intel Core i7-13650HX CPU, 32 GB of RAM, and a NVIDIA GeForce RTX 4060 GPU, which provided robust performance for real-time VR scene generation and interaction. To facilitates information transfer and processing for LVLM, we integrated OpenAI's GPT-4o API. Additionally, a socket-based Python server was established to execute the S-BERT (Sentence-BERT) algorithm, enabling efficient object search and matching during the \textit{Add} operation. In user interactions, users can select objects, manipulate objects, and make UI selections for menus with the ray on right-hand joy stick. Users can open and close menus, select suggestions, and interact with suggestions with the buttons on the left-hand joystick.

\section{Study 1 - Pipeline Evaluation}

We conducted an ablation study to evaluate the impact of different input components in \systemname{}. Considering that the users' verbal instructions for spatial design can vary at abstraction levels and design goals, we designed two additional independent variables in this study. \rewrite{The study has two primary objectives: 
%The study had two primary  objectives:
\begin{itemize}
    \item Evaluate the generation quality of our pipeline components by comparing scene modification quality across four input configurations.
    \item Examine how our pipeline performs for instructions with different abstraction levels.
\end{itemize}
}

\subsection{Study Design}\
To evaluate the effectiveness of the pipeline input information proposed in this paper, we designed this ablation experiment. First, we tested the difference between the final scene results generated by four components conditions containing different information. The four components conditions are:

\begin{itemize}
\item Vision + object parameters + Suggestions $(V+OP+S)$: Includes scene image information (Vision), scene object parameter information (object parameters) and AI-generated suggestions (Suggestions).
\item Vision + Suggestions $(V+S)$: Includes scene image information and AI-generated suggestions.
\item Vision + Object parameters $(V+OP)$: Includes scene image information and scene object parameter information.
\item Object parameters + Suggestions$(OP+S)$: Includes scene object parameter information and AI-generated suggestions.
\end{itemize}

Then, in order to evaluate the impact of modifying scenarios with instructions, we first design our instruction list. Based on the three interior design requirement dimensions: functional requirements, aesthetic style, and psychological stimulus and meaning~\cite{ching2018interior} and three different levels of natural language abstraction (Low, Medium, High)~\cite{sweller2011cognitive, simon1979information}, we designed 9 instructions (3 dimensions × 3 abstraction levels = 9 instructions) as shown in Table~\ref{table_design_commands} and used each instruction to generate with the same initial scenario under four components conditions (36 results, 3 dimensions × 3 abstraction levels × 4 components conditions). Instructions differed between abstraction levels. We counteracted the effects of abstraction level and components conditions. The instructions were carefully crafted to differ across abstraction levels, allowing us to isolate and counterbalance the effects of both abstraction level and components condition on the final scene outcomes.

\begin{table*}[h!]
\centering
\renewcommand{\arraystretch}{1.5} % Adjust line spacing to make it more readable
\begin{tabular}{@{}p{0.20\textwidth}p{0.22\textwidth}p{0.26\textwidth}p{0.26\textwidth}@{}}
\toprule
\textbf{Design goal} & \textbf{Low Abstraction} & \textbf{Medium Abstraction} & \textbf{High Abstraction} \\ \midrule
\textbf{Functional Requirements}  
& \textit{``Add a large screen TV on the wall opposite the couch.''} 
& \textit{``Set up a home theater area for movie nights.''} 
& \textit{``Design a space that brings the cinema experience home.''} \\ \midrule

\textbf{Aesthetic Style}          
& \textit{``Change the sofa color to navy blue.''} 
& \textit{``Apply a nautical theme to the living room.''} 
& \textit{``Evoke the tranquility of the ocean in the living space.''} \\ \midrule

\textbf{Psychological Stimulation} 
& \textit{``Place a small plant on the coffee table.''} 
& \textit{``Introduce elements of nature to enhance relaxation.''} 
& \textit{``Creating a spatial atmosphere that harmonizes with nature to promote balance and relaxation.''} \\ \bottomrule
\end{tabular}
\caption{Natural language instructions tested in Study 1, categorized by design goals and levels of abstraction.}
\label{table_design_commands}
\end{table*}

\begin{table*}[htbp]
\centering
\begin{tabular}{l|l}
\hline
\textbf{Measures}                    & \textbf{Questions}                                                                                             \\ \hline
Relevance                            & How relevant is the scenario generation/modification to the instruction? (Q1)               \\ \hline
Preference                           & How much do you prefer the generation/modification outcome in this condition?  (Q2)          \\ \hline
Reasonableness                       & How reasonable is the scenario generation/modification? (Q3)            \\ \hline
Inspiration        & How inspiring do you find this generation/modification outcome? (Q4)                                                                                             \\ \hline
Open-ended question & Why do you like and dislike the outcome of each condition? (Q5)                                                                                \\ \hline
\end{tabular}
\caption{Questions for Study1 questionnaire.}
\label{Study1Question}
\end{table*}

\subsubsection{\textbf{Task and Procedure}}

The evaluation procedure was designed to enable systematic comparison of scene generation quality across different input conditions. We presented participants with 36 generated scenarios (9 instructions × 4 input configurations) through a standardized slide deck. Each slide displayed an initial scene alongside its modified version generated through one input configuration, with both scenes shown from two distinct viewpoints to provide comprehensive visual context. 

To minimize order effects, we employed a Latin square design to counterbalance both abstraction levels and instructions within each level (3 abstraction levels × 3 instructions = 9 sequences). we implemented full randomization of both instruction sequences and input component condition presentations for each participant. Before beginning the evaluation, participants were asked to review the evaluation guidelines and confirm their understanding of the scoring criteria with the experimenter.

During the study, participants proceeded through the slide deck in their randomized order. For each instruction set, they viewed all four component condition variants (labeled A-D) before providing ratings on a questionnaire. This grouped evaluation approach helped participants make relative judgments across conditions while maintaining the context of each design instruction. After assessing all visual materials for a given instruction, participants recorded their 5-point Likert scale ratings for each dimension and provided qualitative feedback through open-ended responses. 
The complete procedure required approximately 60 minutes per participant.

\subsubsection{\textbf{Data Collection}}
Participants evaluated generated scenes through a structured questionnaire (see Table~\ref{Study1Question}) adapted from a recent work~\cite{Hou2024C2IdeasSC}. The questionnaire assessed four key dimensions below. Each dimension used a 5-point Likert scale (1 = ``Strongly disagre'' to 5 = ``Strongly agree''). 
\begin{itemize}
\item \textbf{Relevance} measured how well the generated scenes matched the instruction intent. Participants evaluated whether elements like object selection and spatial arrangements properly reflected requests.

\item \textbf{Preference} captured subjective satisfaction with the generated outcome in each scene. 

\item \textbf{Reasonableness} assessed physical plausibility, including proper object scaling, absence of collisions, and realistic material properties.

\item \textbf{Inspiration} evaluated the novelty and creative potential of each output, determining whether results sparked new design ideas.
\end{itemize}

Open-ended responses were recorded via audio and later transcribed into text as participants' feedback.

\subsubsection{\textbf{Participants}}

We recruited 18 participants (5 female, 13 male) aged 20 to 30 years (\textit{mean} = 25.5, \textit{SD} = 1.98), recruited through university mailing lists. The participants have diverse academic backgrounds including Engineering, Design/Art, Biology, Chemistry and English. Ten participants reported prior experience with immersive space design applications. 

% We compared the effects of the four components conditions based on the results of the generated/modified scenarios. The questionnaire (as detailed in Table 2) included questions targeting these specific evaluation. Participants were asked to rate their level of agreement with the assessment criteria using a 5-point Likert scale, where 1 indicates "completely disagree" and 5 indicates "completely agree".

\subsection{Results}

We collected a total of 2754 answers (9 instructions × 4 components conditions × 4 categories questions × 18 participants + 9 instructions × 1 open-ended question × 18 participants). Based on these results, we analyzed the data for different components conditions and abstraction levels.

\subsubsection{\textbf{Evaluation of Pipeline Components}}\label{AblationStudyResult}

\begin{figure}%[h]
  \centering
   \includegraphics[width=1\linewidth]{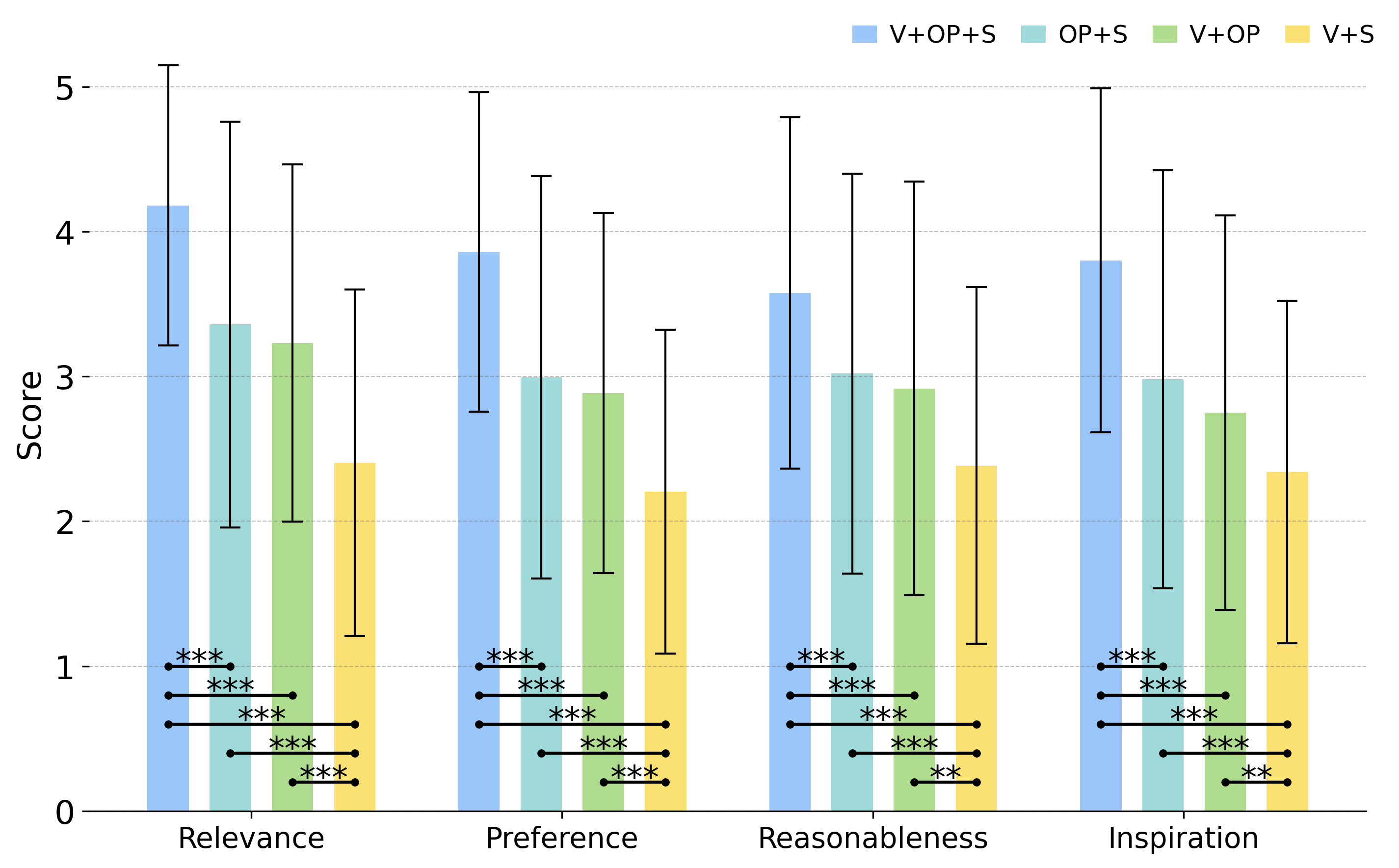}
  \caption{Results of Ablation Study comparing \emph{Input Configurations}. The error bar represents the standard deviation. Statistical significant effects are marked ($*$ = p < 0.05, $**$ = p < 0.01, $***$ = p < 0.001).}
  %\cl{maybe move this one to appendix also? }}
  \label{AblationResult}
\end{figure}

After confirming non-normal distribution, we employed the non-parametric Friedman test (Table~\ref{FriedmanTestComponents}) to detect significant effects in the evaluation results. For pairwise comparisons, we used the Wilcoxon signed rank test. We find that the component condition of \systemname{} ($V+OP+S$) shows the best performance in different categories. As shown in Figure~\ref{AblationResult}, statistically significant comparisons are marked with stars. The evaluations of \emph{Input Configuration} are as follow:

\rewrite{\paragraph{\textbf{Full Pipeline: Vision + SceneInfo + Suggestions($V+OP+S$).}} Our full pipeline outperformed the other three ablated conditions across all the evaluation categories. As shown in Table~\ref{InputMeanTable} and Figure~\ref{AblationResult}, this full input configuration achieved significantly higher scores in relevance, preference, reasonableness, and inspiration than all the other configurations. These results demonstrate that each of the three input components plays an essential role in the system's performance.} 

\begin{figure*}[htbp]   
\centering
\includegraphics[width=1\textwidth]{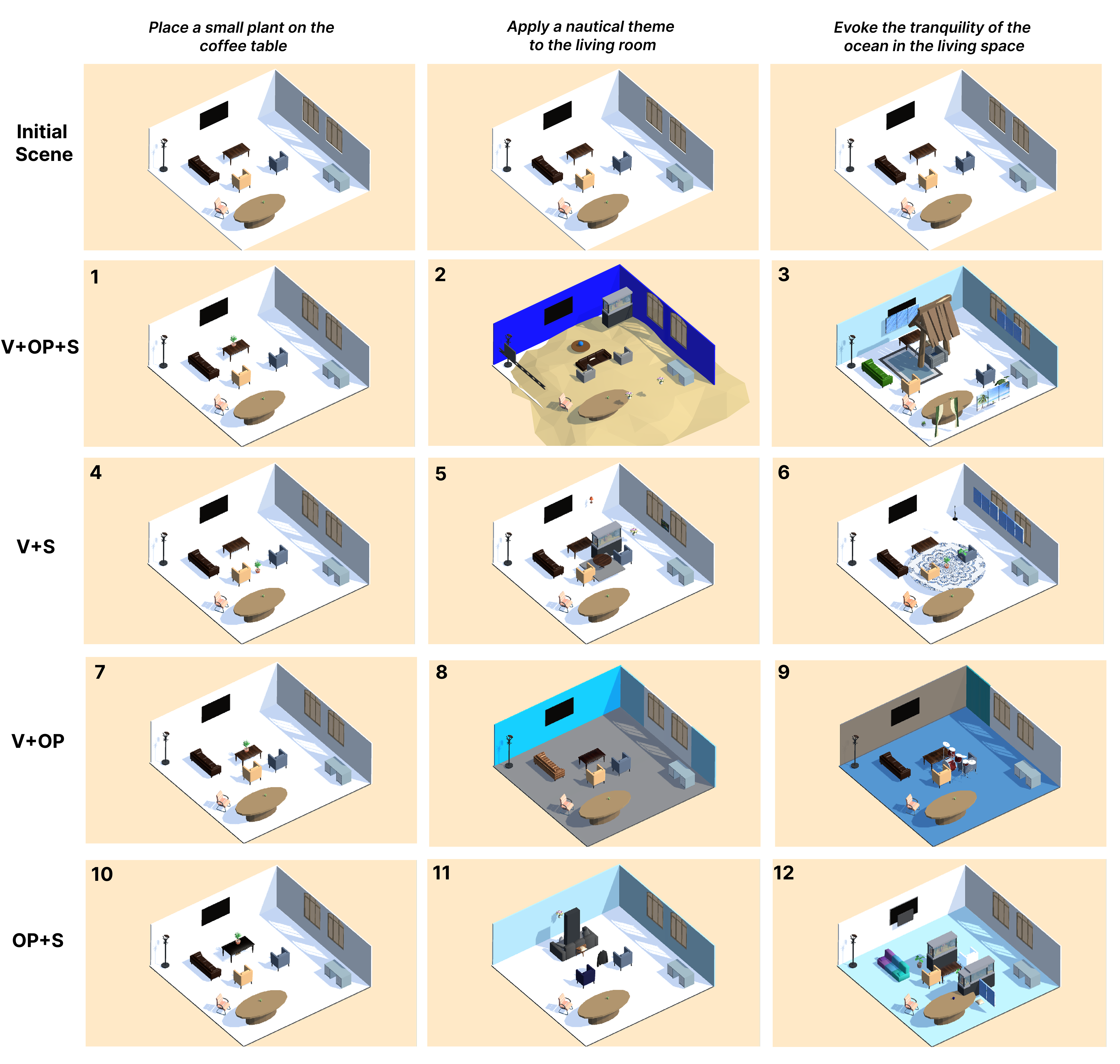}
	
	\caption{Results of scene modification under four pipeline input configuration conditions. The first row is the starting VR scene for all conditions, each subsequent row is for one condition. The columns from left to write show examples of instructions in low, medium to high abstraction level.}
    \label{AblastionStudyPictures}
\end{figure*}

% The integration of visual input, object parameters, and AI-generated suggestions enables \systemname{} to generate outputs that are not only accurate and contextually appropriate but also diverse and creatively inspiring. 
\rewrite{From open-ended questions in the questionnaires, we found this input configuration improved the ability of \systemname{} to generate thematically appropriate content with high relevance, such as beach and beach volleyball (Figure~\ref{AblastionStudyPictures} (2)) and executing instructions with minimal deviation (Figure~\ref{AblastionStudyPictures} (1)). It maintained spatial accuracy, with correctly placed and oriented furniture, and achieved stylistic consistency, rendering elements like nautical wallpapers and ocean-themed decorations that aligned with user-specified atmospheres (Figures~\ref{AblastionStudyPictures} (2–3)). Finally, \systemname{} exhibited strong inspirational potential, introducing unexpected yet fitting elements—such as sand or a well—that enriched the scene and exceeded user expectations without sacrificing contextual fidelity (Figure~\ref{AblastionStudyPictures} (2)).} 

\rewrite{\paragraph{\textbf{Removing Object Parameters ($V+S$).}} This input configuration performed significantly worse than all the other conditions across all evaluation categories. This condition lacks object parameters, which likely undermines the ability of AI to interpret spatial context and generate coherent or appropriate modifications. Without access to parameters such as position and scale, AI struggles to produce relevant, well-aligned, or inspiring outcomes, despite having visual input and suggestions. The system struggled to identify relevant objects, resulting in poor thematic alignment. As shown in Figure~\ref{AblastionStudyPictures} (5)--(6), objects in the scene are not coherent to the ocean theme. Layout errors are more severe, such as furniture overlap in Figure~\ref{AblastionStudyPictures} (5)--(6) and illogical spatial arrangements.}

\rewrite{\paragraph{\textbf{Removing Vision ($OP+S$) and removing suggestions ($V+OP$).}} These two conditions had mediocre performances compared with other conditions. While $OP+S$ yielded slightly higher mean scores than $V+OP$ in all evaluation categories, the differences were not statistically significant. %This suggests that neither visual input nor suggestions alone can compensate for the absence of the other, and that the synergy between all three components is necessary for producing high-quality, user-aligned scene modifications. 
From the open-ended questions, we found that without vision ($OP+S$), it impaired the LVLM's spatial reasoning and color coordination. This was reflected in severe layout issues, including overlapping objects and irrational layouts such as the sofa overlaps with bookshelves in Figure~\ref{AblastionStudyPictures} (11). Color perception suffered similarly, for ocean-themed modifications, results were reported as ``not blue enough'' in Figure~\ref{AblastionStudyPictures} (7). On the other hand, the absence of suggestions ($V+OP$) constrained the LLM's creative capacity, producing minimal modifications, such as Figure~\ref{AblastionStudyPictures} (8). It also frequently introduced contextually inappropriate objects like a drum kit in a ocean themed room in Figure~\ref{AblastionStudyPictures} (9).} 

\begin{figure}%[htbp]%[h]
  \centering
   \includegraphics[width=1\linewidth]{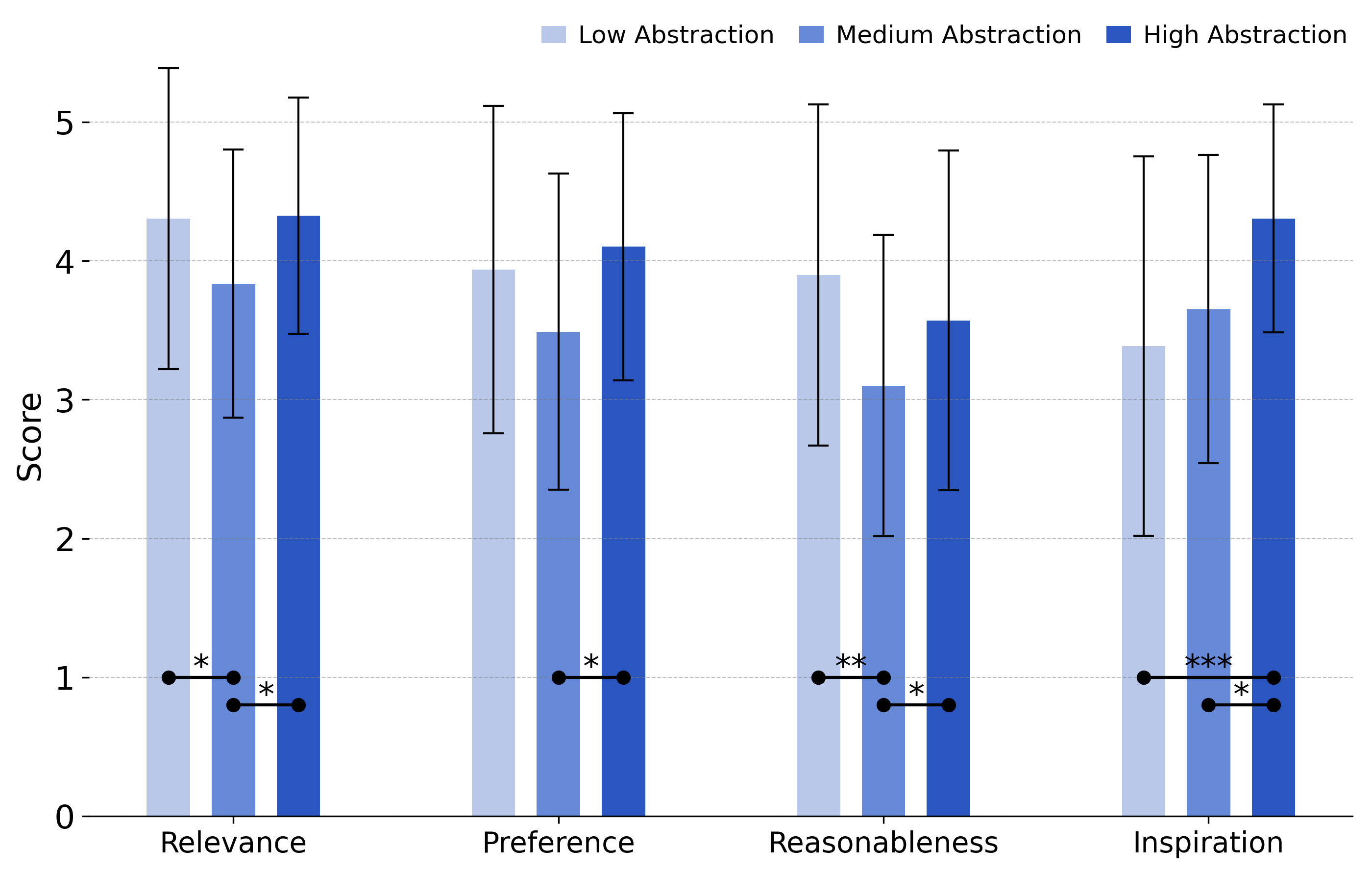}
  \caption{Results of comparing \emph{abstraction levels} (only in $V+OP+S$ - full pipeline condition). The error bar represents the standard deviation. Statistical significant effects are marked ($*$ = p < 0.05, $**$ = p < 0.01, $***$ = p < 0.001).}
  \label{AbstractionLevles}
\end{figure}

\subsubsection{\textbf{Effects of Abstraction Levels}}
We analyzed the effects of \emph{Abstraction Level} on the generated results using only the data in the full pipeline condition $(V + OP + S)$ (\systemname{}) across levels of abstraction. 
Same as section~\ref{AblationStudyResult}, we confirmed non-normal distribution of data, employed Friedman test to detect significant effects. For pairwise comparisons the Wilcoxon signed-rank test was used. \rewrite{While details of the statistical analysis results are in the Appendix A.3, we highlight the main findings here.}

\rewrite{Overall we can see in Figure~\ref{AbstractionLevles}, the system performed better for instructions at low and high abstraction level than at medium abstraction for Relevance, Preference and Reasonableness, but not for Inspiration. Instructions at a higher abstraction level seem to generate more inspiring outcome.}

\rewrite{\paragraph{\textit{\textbf{Low Abstraction}}}
% Our pipeline performed similarly well in \emph{Relevance \& Reasonableness} for low abstraction instructions. 
We can see that low abstraction instructions shows high relevance score and reasonableness score which are significantly higher than medium abstraction instructions. This might be caused by their explicit nature, yielding precise executions that from open-ended questions answers called ``precise'' and ``directional''. However, this specificity came at a cost: while functionally reliable, this condition scored poorly in \emph{Inspiration} which is significantly lower than other abstraction levels, as the predictable outcomes of low abstraction instructions don't need additional operations, which offers little novelty.}

\rewrite{\paragraph{\textit{\textbf{Medium Abstraction}}}
Our pipeline performed worse for medium abstraction instructions across metrics. For \emph{Relevance \& Reasonableness}, significantly lower than low abstraction and high abstraction (Figure~\ref{AbstractionLevles}). Medium abstraction instructions occupied an awkward middle ground — clearer than high abstraction but vaguer than low abstraction. In this condition, we find answers of open-ended question were more sensitive to imperfections, such as one extra object will make the answer negative. This probably impacted \emph{Preference}, where this condition was least favored due to perceived lack of creativity and inconsistent execution (Figure~\ref{AblastionStudyPictures} (2), the sand is a little strange). Its \emph{Inspiration} score suffered similarly, with outputs deemed neither reliably accurate nor meaningfully novel, failing to deliver on the strengths of either extreme.}

\rewrite{\paragraph{\textit{\textbf{High Abstraction}}}
Our pipeline performed strongly across all metrics for high abstraction instructions. For \emph{Relevance \& Reasonableness}, the score of high abstraction is significantly higher than medium abstraction instructions in all categories and higher than low abstraction instructions in  \emph{Inspiration}. From the answers of open-ended questions, they valued the balance between thematic coherence and interpretive freedom, accepting minor inconsistencies when the overall atmosphere aligned with their vision (e.g., ``extra content but the overall atmosphere was good''). In terms of \emph{Preference}, this level was favored for enabling creative diversity, with even imperfect layouts often perceived as intentionally artistic. This condition outperformed both other abstraction levels in \emph{Inspiration}, generating outputs described as ``exciting'' and ``fresh'', as the LLM's broad interpretations often surprised and delighted users.}

\section{Study 2 - User evaluation}
The goal of this study is to evaluate the effectiveness of displaying AI-generated interactive suggestions following users' natural language instructions. We created a baseline condition to compare with \systemname{} for this purpose. The baseline is an automated VR scene modification approach based on the same backend pipeline of \systemname{}. \rewrite{It does not provide interactive suggestions, but directly modifies the scene following users' verbal instruction. To keep the final generation quality consistent across conditions, we kept the ``suggestions generation'' module in the baseline backend although suggestions are not displayed users.} %Under this condition, participants input instructions, and the system directly generates or modifies VR scene without showing intermediate AI suggestions. 

The study aimed to address two core research questions: 

% \rewrite{\textbf{RQ1.} How do users engage with \systemname{} and the ETE approach for progressive scene design? }
\rewrite{\textbf{RQ1.} How does \systemname{} compare with the baseline in terms of user satisfaction, workload and experience?} 

\rewrite{\textbf{RQ2.} What creative strategies and workflows do participants adopt when using \systemname{} and the baseline, respectively?}
% (RQ1) How do users engage with suggestion-based interfaces and the ETE approach for progressive scene design? 

% \textbf{RQ3.} How does seeing and interacting with suggestions affect user experience and creative outcomes compared to opaque generation?

\subsection{Design}

We designed a within-subjects experiment to compare \systemname{} and \rewrite{the baseline}. 

\subsubsection{\textbf{Participants}}
We recruited 12 participants (6 female, 6 male) aged 23-29 (\textit{M}=25.7, \textit{SD}=1.65) through university mailing lists and local VR interest groups. Screening ensured all participants had limited professional 3D design experience (<1 year) but were familiar with VR interfaces (10/12 reported regular HMD use). This profile represented our target user base of non-expert designers who might benefit from AI assistance. Participants received \$20 compensation for the approximately 2-hour session, including training, tasks, and interviews.

\subsubsection{\textbf{Aparatus}}
The experiment used Unity 3D on Meta Quest 2 headsets, with researchers observing via a tethered laptop connection. Sessions were video recorded with participant consent.

\subsubsection{\textbf{Task}}

The same as Study 1, we developed three \rewrite{task types} that reflected key design goal dimensions: functional, aesthetic and psychological stimulation. 
The study followed a structured protocol to ensure consistent data collection:
\begin{itemize}
\item \textbf{Functional Requirements:} Designing a living room/bed-

room/study room with a XXX functional requirement.
\item \textbf{Aesthetic Style:} Designing a living room/bedroom/study room with a XXX aesthetic style.
\item \textbf{Psychological Stimulus and Meaning:} Designing a living room/bedroom/study room with a XXX psychological stimulus and meaning.
\end{itemize}

\subsubsection{\textbf{Procedure}}

To begin with, participants completed a training session by first watching a system introduction video covering \systemname{} and baseline interfaces as well as manual scene editing controls. They were briefed on the think-aloud protocols and study procedure. They were asked to perform training tasks that familiarized them with suggestion interaction and regeneration features.
    
In the main session, each participant completed two tasks, each of a different type, using both \systemname{} and baseline in a counterbalanced order. To reduce fatigue, participants were randomly assigned two out of the three available task types. This resulted in a total of four tasks per participant (2 types × 2 conditions). The tasks involved designing rooms with $\ge$ 10 objects, starting from an identical empty VR space. 
    
After each task, participants completed the NASA-TLX questionnaire~\cite{hart1988development} to measure perceived mental, physical, and temporal demands. Participants rated both conditions on 7-point Likert scales across four dimensions (preference, inspiration, control, implementation) and participated in a semi-structured interview about their experiences.

\subsubsection{\textbf{Data Collection and Analysis}}
We employed a mixed-method approach to capture both behavioral and subjective measures:

\paragraph{\textbf{Behavioral Data}}
System logs recorded timestamped interactions including: voice commands, suggestion applications/regenerations, manual edits (additions, deletions, transformations), and undo operations. These were synchronized with think-aloud audio and screen recordings for contextual analysis. We summarized patterns in design and operation strategies by triangulating behavioral logs and think-aloud observation. 

\paragraph{\textbf{Quantitative Measures}}
The post-task questionnaire assessed \emph{User Satisfaction} using four 7-point Likert scales (1=low, 7=high) for user preference, inspiration, user control, and system execution. NASA-TLX scores measured cognitive load across six subscales (mental, physical, temporal demands; performance; effort; frustration).

\paragraph{\textbf{Interview data}}
Two researchers independently coded 25\% of the interview transcripts using thematic analysis and achieved agreement. Discrepancies were resolved through discussion to finalize the themes.

\subsection{Results}

\rewrite{In this section, we present both quantitative and qualitative findings. The quantitative analysis is based on questionnaire ratings, while the qualitative insights are derived from behavior logs, think-aloud sessions, and interview transcripts. The think-aloud and interview data were analyzed using thematic analysis. Two researchers first reviewed the data from two participants to establish an observational protocol and identify initial themes. Once consensus was reached, one researcher proceeded to analyze the full dataset. We organize our findings based on our research questions in the following. }

% \rewrite{Firstly, to answer RQ1, we introduce the results of comparing our system with the ETE condition in Sections 5.2.1. Next, we answer the RQ2 in section 5.2.2. Finally, we present user-suggested improvementsin section 5.2.5.}

% We report both quantitative and qualitative findings in this section. Quantitative findings come from ratings from the questionnaires.
% Qualitative findings come from observing the behavior logs, user think-aloud sessions, and interview transcripts. Think aloud and interview transcripts were analyzed through thematic analysis. Two researchers went through the data of two participants and agreed on the observational protocol and identified themes before one of proceeded to analyze the entire data set. We report patterns in users' strategies and operations when crafting VR scenes using our system versus the ETE condition.

\subsubsection{\textbf{User Satisfaction, Experience and Workload (RQ1)}} %with \systemname{} and baseline (RQ1)}}

\begin{figure}%[h]
  \centering
   \includegraphics[width=1\linewidth]{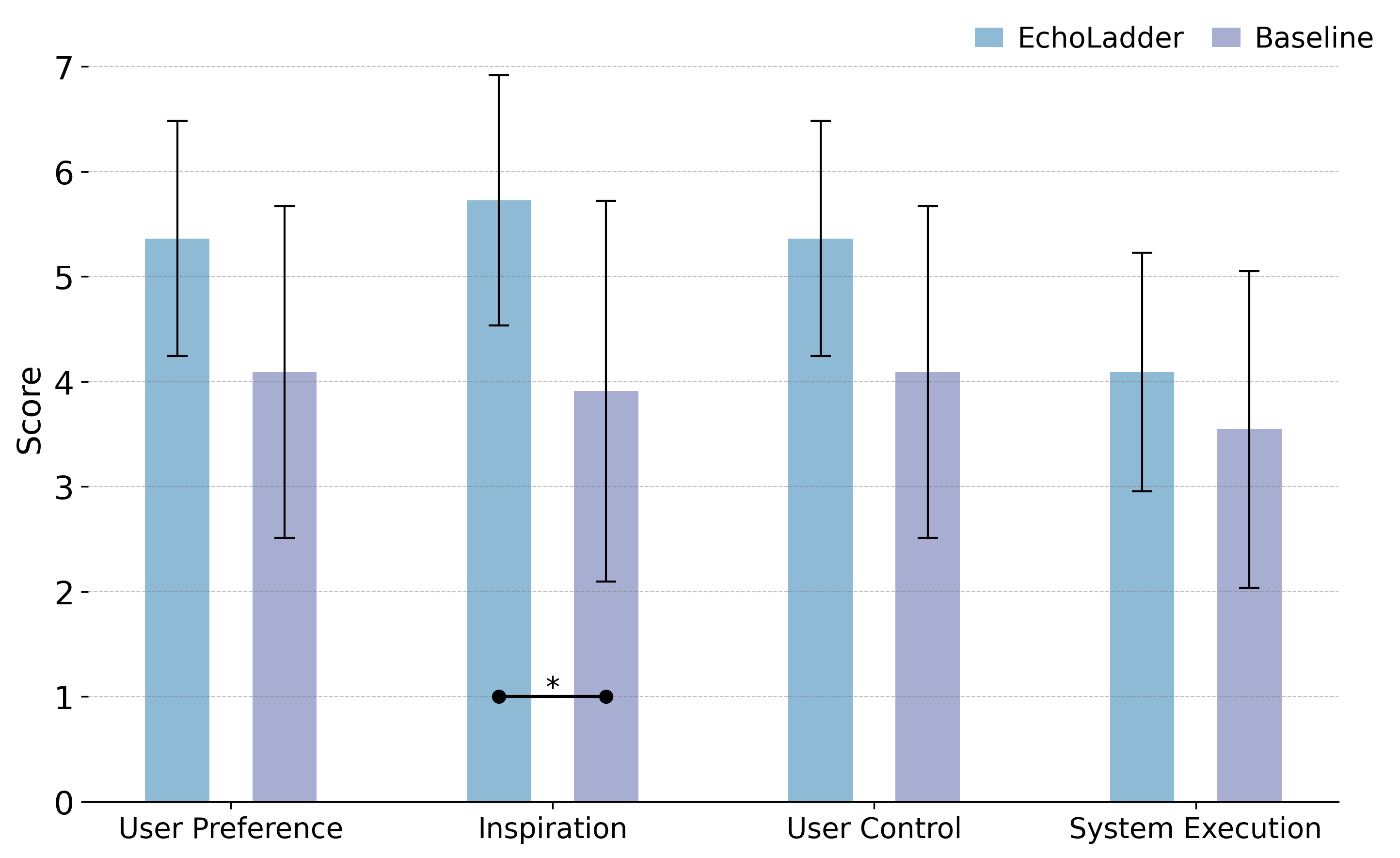}
  \caption{\rewrite{Subjective ratings from participants for EchoLadder vs. Baseline.} The error bar represents the standard deviation. Statistically significant effects are marked ($*$ = p < 0.05).}. 
  \label{SurveyScore}
\end{figure}

\paragraph{\rewrite{\textit{\textbf{User Satisfaction, Control and Engagement.}}}} \rewrite{Based on user subjective ratings (as shown in Figure~\ref{SurveyScore}), \systemname{} achieved significantly higher Inspiration scores than baseline (mean = 5.75 vs. 4, p = 0.032), indicating stronger creativity support. Its average rating also outperformed the baseline on user preference, perceived user control and quality of suggestion execution.}

% \paragraph{\textit{\rewrite{Sense of Control and User Engagement}}}
In the interviews, 7 out of 12 participants agreed that \systemname{} fostered a stronger sense of agency in decision-making and content modification due to the flexibility of accepting, rejecting, or adjusting AI-generated suggestions. As P2 noted, \systemname{} felt \textit{``more reassuring''} because it allowes \textit{``make independent decisions and modify every AI-proposed change''}. Specifically, 6 participants appreciated seeing the AI’s reasoning process through textual suggestions, which enhanced controllability through procedural visibility. As P9 stated: \textit{``\systemname{} makes me feel like I’m a grading teacher and can see and interact with AI's throught process.''} Additionally, 7 participants valued the ability to apply, undo, or regenerate suggestions, granting them operational flexibility and modification authority at each stage. For instance, P11 mentioned: \textit{``Because suggestions can be undone and regenerated multiple times, I can try them one by one. If I don’t like them, I can just stop generating''}. P8 added that most of the time LLMs fail to fully understand the intention so the option of selective generation allows them to \textit{``cut losses midway''}. Overall, \systemname{} fosters stronger control and engagement and make user feel like \textit{``I am the master, and it is just a tool''} (P9). 

% Log analysis, as illustrated in Figure~\ref{ManualCountRate}, further revealed \systemname{} required significantly fewer manual operations (mean = 0.67 vs. 0.79, p = 0.003), suggesting reduced effort on trivial tasks or greater engagement in hands-on crafting.}

\paragraph{\rewrite{\textit{\textbf{Perceived Work Load and Manual Operations}}}} %\jz{Compress this paragraph in one sentence? there is no significance} 

In terms of perceived workload measured through NASA TLX (Figure~\ref{NASATLX}), we ran a Wilcoxon signed-rank test due to a violation of normality. We found no statistical difference between \systemname{} and the baseline on any of the measures. %\todelete{We report the means and SDs for each condition as follows. Mental Demand (\systemname{}: $mean = 11.50, SD = 5.02$, ETE: $mean = 9.58, SD=5.18$), Physical Demand (\systemname{}: $mean = 9.96, SD = 5.20$, ETE: $mean = 9.79, SD=4.67$), Temporal Demand (\systemname{}: $mean = 9.88, SD = 4.53$, ETE: $mean = 9.63, SD=4.99$) and Effort (systemname{}: $mean = 11.75, SD = 4.32$, ETE: $mean = 10.54, SD=5.59$). Performance (systemname{}: $mean = 9.46, SD = 4.85$, ETE: $mean = 9.5, SD=5.49$) and Frustration (systemname{}: $mean = 8.13, SD = 4.27$, ETE: $mean = 8.38, SD=5.95$). Our interview findings could help explain the mixed ratings for mental and physical load. } 
\rewrite{Qualitative findings revealed some potential causes of mental and physical load. Regarding mental load, while many appreciated \systemname{}'s structured approach, some found it taxing: P1 compared it to ``\textit{converting word problems into multiple-choice questions},'' and others (P2, P3) experienced decision fatigue from evaluating numerous suggestions. P3, a non-designer, felt it was ``\textit{more demanding}'' and preferred the baseline's ready-made results. However, the baseline's ``\textit{all-at-once}'' generation (P7, P9) often overwhelmed users when outputs missed their intent, sometimes discouraging iteration—as seen in P4's passive ``\textit{Just like it}'' acceptance. In terms of physical load, both conditions required similar effort for fine-tuning, but the baseline demanded more corrective actions, with users frequently deleting misplaced objects (P1, P6, P11). \systemname{} reduced ``\textit{sunk effort}'' (P9) by allowing early rejection.}

\begin{figure}%[htbp]
\centering
 \includegraphics[width=0.5\textwidth]{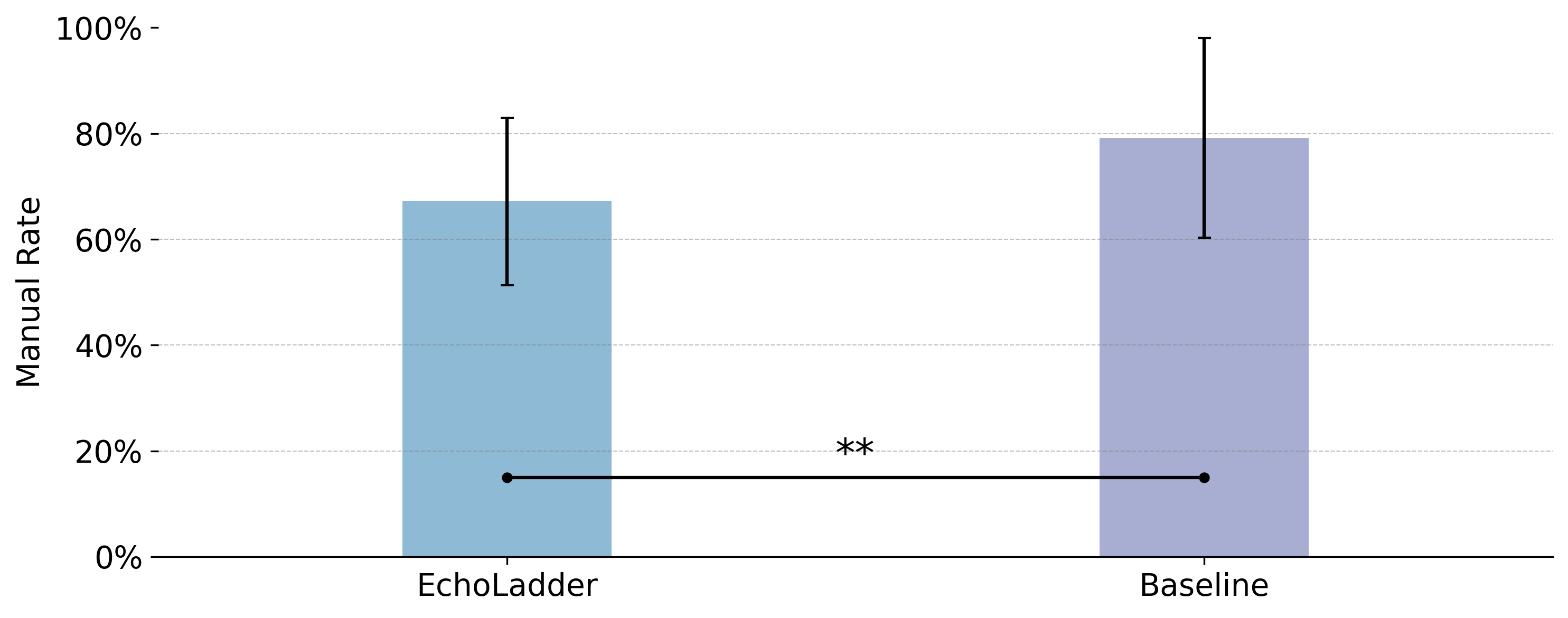}
\caption{Average percentage of manual operations using \systemname{} and baseline. Statistical significant effects are marked ($**$ = p < 0.01).}
  \label{ManualCountRate}
\end{figure}

\rewrite{To provide an additional perspective on workload,} we computed the ratios of manual operations in each task from all participants' interaction logs. We examined the normality of the data and tested for significance by t-test. As illustrated in Figure~\ref{ManualCountRate}, the manual operation rate of \systemname{} (\systemname{}: $mean = 0.6715, SD = 0.159$, Baseline: $mean = 0.7916,SD = 0.189$, $p = 0.003$) was significantly lower in Possible interpretations include \systemname{} saved more effort of trivial manual operations, and/or participants engaged more with hands-on crafting in the design process.

\begin{figure}%[h]
  \centering
   \includegraphics[width=1\linewidth]{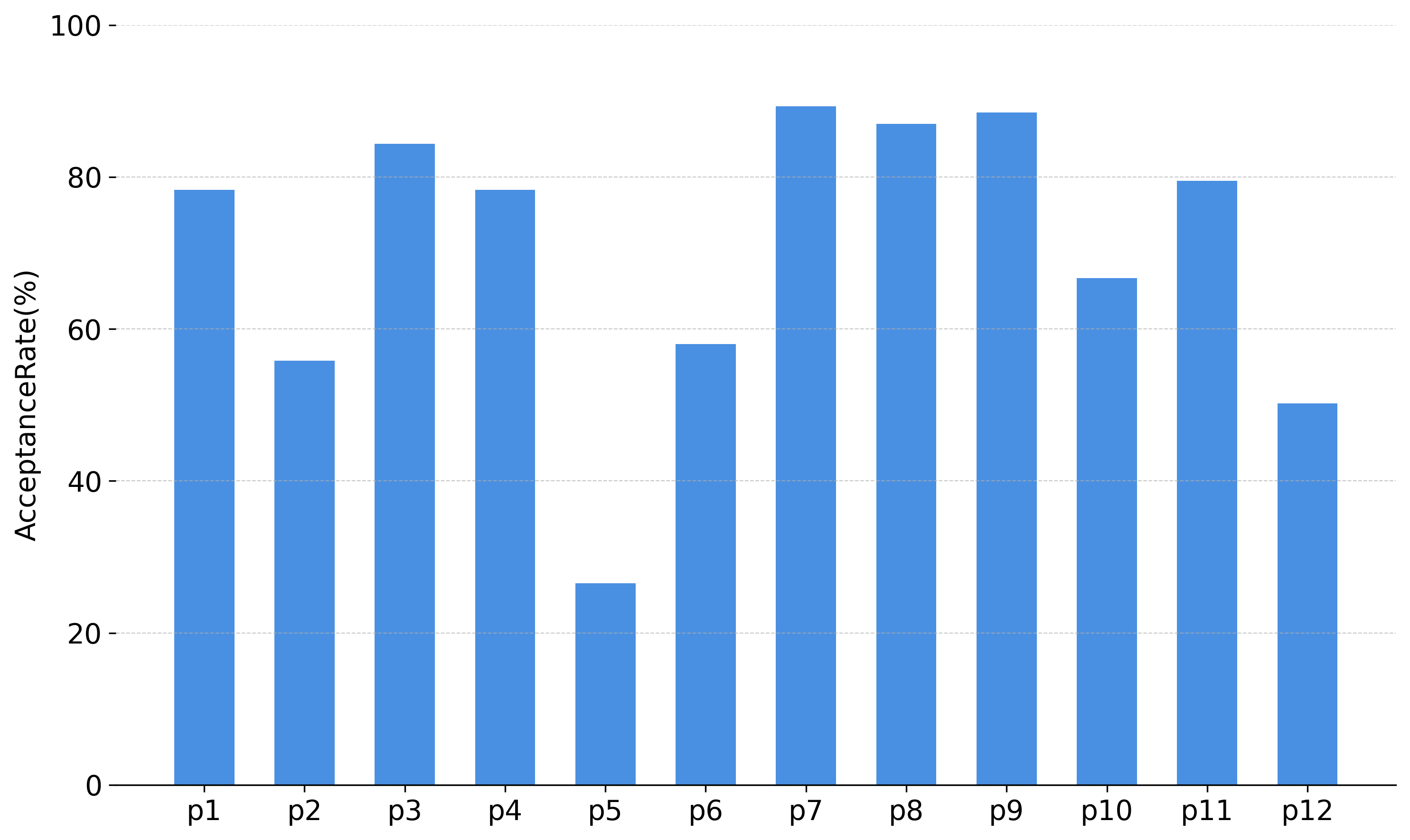}
  \caption{Average suggestions acceptance rate of each participant.}
  \label{AcceptanceRateAverage}
\end{figure}

% \subsubsection{\textit{\rewrite{The Effects of Interactive Suggestions}}}
\paragraph{\textit{\rewrite{\textbf{Effects of Providing Suggestions in EchoLadder}}}}

% \paragraph{\textit{\rewrite{Suggestion Acceptance rate in \systemname{}}}}
\rewrite{As providing interactive suggestions is the core feature of our system's interface innovation, we dived into analyzing how participants used it and understanding its effects. }

\rewrite{We first calculated the \emph{Suggestion Acceptance Rate} from the system log as the ratio of accepted suggestions (applied and not retracted) to the total suggestions provided per task. The overall acceptance rate averaged 70.2\% (SD = 0.196), with 7 of 12 participants exceeding the mean, though individual rates varied widely (e.g., 90\% for P7 vs. 23\% for P5), as shown in Figure~\ref{AcceptanceRateAverage}. This indicates that our system-generated suggestions are generally reasonable and well-used. Three main reasons for rejecting suggestions are identified from interviews:} (1) \emph{Content mismatches}, where proposals conflicted with design intent (e.g., P12 dismissing a dining table—\textit{``I don't want this''}, Style Task S1); (2) \emph{Poor execution} of accepted suggestions, later undone due to unsuitable colors (P2/P8) or implausible layouts (\textit{``This implementation is crazy''}, P2);
and (3) \emph{Redundancy}, where users ignored excess suggestions after core needs were met (P12 stopping after four proposals: \textit{``Content is sufficient—I'll adjust manually''}).

% \subsubsection{\textbf{Suggestion Acceptance Rate in \systemname{}}}

% From the system log, we calculated \emph{Suggestion Acceptance} based on whether a suggestion was applied to modify the scene and not retracted. Participants usually issued multiple verbal instructions to the system and iteratively refined the scene. The suggestion acceptance rate for each task is calculated as the total number of accepted suggestions throughout the task divided by the total number of suggestions provided. 

% Figure~\ref{AcceptanceRateAverage} reports the overall suggestion acceptance rate for each participant ($mean = 70.2\%, SD = 0.196$). 7 out of 12 participants
% demonstrated above-average suggestion acceptance rates, while individual variances remain: eg. P7 accepted almost 90\% of the suggestions and P5 only accepted 23\%. 

% Our interview identified three main reasons for suggestions being rejected by our participants. Participants rejected suggestions for three key reasons: (1) \emph{Content mismatches}, where proposals conflicted with design intent (e.g., P12 dismissing a dining table—\textit{"I don't want this"}, Style Task S1); (2) \emph{Poor execution} of accepted suggestions, later undone due to unsuitable colors (P2/P8) or implausible layouts (\textit{"This implementation is crazy"}, P2);
% and (3) \emph{Redundancy}, where users ignored excess suggestions after core needs were met (P12 stopping after four proposals: \textit{"Content is sufficient—I'll adjust manually"}).

% \paragraph{\textit{\rewrite{Three Advantages of Showing Suggestions}}}
Interview findings reflected three key benefits of the suggestion-based approach:

\begin{enumerate}
    \item \textit{Decision Support \& Creativity Stimulation}. Suggestions effectively scaffolded the design process by addressing both ideation bottlenecks and choice overload. For users struggling with initial direction (e.g., P2's uncertainty about color schemes or P12 \textit{``didn't know what to do next''} shown in Figure~\ref{Workflow} Task 2), suggestions provided concrete starting points. Conversely, when overwhelmed by options, participants like P4 appreciated how suggestions helped \textit{``narrow thinking directions''}. The sequential nature of suggestions also fostered creative connections—P7's experience typified this, where an initial sofa color suggestion naturally led to complementary furniture adjustments, creating design coherence.
    \item \textit{Spatial Awareness \& Control}. Suggestions enhanced 3D scene manipulation through explicit spatial references. As P6 noted, directional cues helped users ``accurately locate new elements'' in the immersive environment. The modular workflow allowed incremental adjustments that participants found more manageable than holistic generation, e.g.,\textit{``Modifying elements one-by-one gives better control''} (P7). However, this precision came at a cost: P7 and others reported \textit{``higher workload from individual adjustments''}, indicating a trade-off between control and efficiency.
    \item \textit{Design Intent Preservation}. Unlike baseline's monolithic generation, suggestions maintained stronger alignment with user intentions through stepwise refinement. When initial proposals missed the mark (e.g., P8's Chinese lantern suggestion), follow-up recommendations (e.g., matching bookshelf ornaments) enabled course-correction while preserving the overall design vision. This iterative process reduced the \textit{``start-from-scratch''} frustration observed in baseline's workflows.
\end{enumerate}

\rewrite{\systemname{}'s \textit{Undo} and \textit{Regenerate} functions proved critical for managing AI-generated content, with the \textit{Undo} feature used by 9 out of 12 participants to efficiently correct mismatches between suggestions and design intent, such as removing unsuitable furniture (P2) or reverting unwanted color changes (P5).} Moreover, six participants also employed \textit{Undo} as a diagnostic tool to detect scene changes outside their HMD's field of view, which is a known issue in 3D virtual environments~\cite{gruenefeld2019comparing}. The \textit{Apply–Undo–Apply} strategy (e.g. Figure~\ref{Workflow} P2 Task 1 \& P12 Task 3) helped identify subtle modifications, addressing cases where changes were initially imperceptible or too minor to notice (P7,P8). This contrasted with the baseline, where P7 noted difficulty in tracking changes. %: \textit{``Does this method add something? I don't see anything.''}

\subsubsection{\rewrite{\textbf{Creative Strategies and Workflows in the Two Conditions (RQ2)}}}

Our analysis revealed systematic differences in how participants approached scene design across the two conditions. These manifested in both design iteration activities and operational workflow patterns. Both conditions shared the same core design activities and showed slight differences in their iterative workflows and operational patterns. We also found similarities and differences in their creativity support, as reported below.

\paragraph{\textbf{Common Design Activities}}
Participants engaged in three fundamental design activities regardless of the condition. \textit{Global planning} involved high-level conceptualization of space functions and aesthetics, as exemplified by P2's comprehensive vision: \textit{``First, I established this should be a war-themed bedroom with military decor''}. This typically preceded \textit{targeted modifications} of specific aspects—P6's focused adjustment (\textit{``Now make the walls camouflage green''}) being characteristic. Finally, all participants performed \textit{object-level manipulations}, though with differing frequency; P8's precise placement (\textit{``The TV needs 30cm clearance from the couch''}) typified this granular control.

\paragraph{\textbf{Slightly Different Iterative Workflows}}
Building upon these design activities, we identified three composite workflow strategies.
The most common was \textit{top-down refinement}, where participants like P4 progressed systematically from global concepts to specific implementations as shown in Figure~\ref{Workflow} P4 Task 1: \textit{``I first defined a 'relaxing lounge' concept, then selected appropriate furniture styles, and finally adjusted individual pieces''}. This contrasted with the \textit{focused execution} approach favored by P6 and others, who transitioned rapidly to object manipulation after minimal planning (\textit{``I knew it needed a TV, so I placed it first and built around it''}).

Interestingly, the baseline enabled a unique \textit{cyclic refinement} pattern absent in \systemname{}. As P9 described: \textit{``I kept oscillating between adding artworks and tweaking their arrangements—each adjustment inspired new ideas''}. This back-and-forth process was observed in three participants, suggesting a more exploratory nature of using the baseline to design, in contrast to a more top-down approach with \systemname{}.

\paragraph{\textbf{Different Operational Patterns}}
The systems elicited different operation patterns. \systemname{} users predominantly adopted either \textit{sequential} (n=8) or \textit{batch} (n=4) processing. Sequential users like P7 emphasized control: \textit{``Applying suggestions one-by-one lets me catch issues early''}. Batch processors like P4 valued efficiency: \textit{``I execute everything first, then clean up—it's faster overall''}.
The baseline, by contrast, enabled three distinct modes. The \textit{asynchronous} approach was observed in participants (n=5) like P6 (Figure~\ref{Workflow} Task 1) multitasking during generation: \textit{``While the AI worked on walls, I placed furniture''}. Others (n=7) preferred \textit{post-generation review}, with P2 (Figure~\ref{Workflow} Task 2 \& 3) noting: \textit{``I let the AI finish completely before making any edits''}. In addition, two \textit{manual-centric} participants manually established bases before AI involvement. P9's explanation is: \textit{``I needed to 'anchor' my vision first''}.

\paragraph{\rewrite{\textbf{Creativity Support}}}
\rewrite{Three types of creativity support emerged in both conditions.} First, the system \emph{clarified vague ideas} by materializing abstract concepts. Participants reported sudden clarity when seeing concrete suggestions, with P4 noting: \textit{``The sofa and bookshelf suggestions made me instantly visualize arrangements''}. This effect was particularly strong for users with limited initial vision—P7, who struggled with sports-themed designs, found the treadmill suggestion pivotal, while P4 described how basic items like beds naturally prompted complementary additions (\textit{``A rug underneath came to mind immediately''}).
Second, AI proposals \emph{broadened design considerations} by surfacing overlooked elements. P6's experience was typical: \textit{``The system reminded me about lighting and contrast—aspects I'd neglected''}. This expansion occurred both for functional properties (visibility, spatial relationships) and aesthetic dimensions (color coordination, stylistic coherence).
Third, users frequently \emph{adopted unexpectedly fitting suggestions} that diverged from their initial plans. P6's incorporation of a chandelier in a princess-themed room exemplified this: \textit{``I hadn't considered how crystal lights would perfect the Barbie aesthetic''}. Such discoveries often produced what P11 described as \textit{``a sudden spark of inspiration''}.

\rewrite{The two conditions also supported creativity differently in some aspects.} \systemname{}'s textual suggestion lists provided conceptual starting points, while the baseline generation's concrete visuals stimulated more immediate reactions. P9 articulated the difference between the two modes: \textit{``Text is abstract, but objects are tangible. With the baseline, I still need to mentally process what I see — but \systemname{} directly provides the thought process''}. This might explain the higher rating on Inspiration in the survey result (Section 5.5.1). However, this potential came with variability—P7 rejected an unwanted traditional Chinese style scene, but valued the \textit{regenerate} option for managing unpredictability: \textit{``Quick regeneration makes the AI's randomness feel controllable''}.

\subsubsection{\textbf{User-suggested Improvements}} We collected the following user suggestions during their think-aloud and interview answers.

\paragraph{\textbf{Selective scene iteration}} Currently, \systemname{}’s AI considers all objects in the scene when generating suggestions, which lacks the flexibility of allowing only specific objects to participate in iterative AI suggestions or targeting individual objects for refinement. On one hand, participants emphasized the need to preserve manually adjusted elements from further AI iterations. For instance, P1 stressed the importance of keeping \textit{``my completed operations unaffected''} and having \textit{``the option to selectively include elements for AI modification''}. Similarly, P1 and P7 expressed frustration when AI suggestions altered their manually optimized objects. On the other hand, P1 requested a \textit{``focus mode''} for modifying individual objects without affecting others, noting: \textit{``I need the ability to change the style of a single object independently without influencing other objects''}.

\paragraph{\textbf{Iterative prompt crafting}} Participants expected the ability to refine prompts iteratively. P1 wanted to \textit{``iterate a prompt by modifying the current one''}, while P3 mentioned the need to \textit{``regenerate a single suggestion if the output is unexpected''}.

\paragraph{\textbf{Multimodal prompts}} P1 highlighted the need for multimodal prompts, stating: \textit{``I’d prefer prompts to include visual sketches alongside text''}.

\paragraph{\textbf{Beyond first-person perspective}} Participants identified challenges in spatial orientation within virtual environments. P7 noted, \textit{``Operating on objects one by one in a first-person 3D space is fatiguing''}, while P5 remarked: \textit{``I struggle to gauge the scale of the entire scene and would greatly benefit from a god’s-eye view to navigate and inspect the space from above''}. These insights suggest a need for improved scene visualization tools to support navigation and decision-making.

\section{Discussion}

\subsection{Summary of findings}

Our evaluation of \systemname{} demonstrates its effectiveness in enabling progressive, user-guided VR scene design through AI-generated interactive suggestions. The ablation study confirmed the necessity of integrating visual input, object parameters, and AI suggestions (V+OP+S) to achieve optimal scene generation quality. This configuration outperformed ablated variants in relevance, reasonableness, and inspiration, particularly for high-abstraction instructions, where the system performed well at translating abstract concepts into coherent spatial designs. 

The user study revealed that \systemname{} significantly enhanced user creativity and inspiration compared to the baseline generation, with participants leveraging suggestions to iteratively refine their designs while retaining agency. By providing textual suggestions, \systemname{} offers figurative hints that trigger spatial associations and conceptual thinking, introducing unexpected inspiring elements. Key interaction features—undo, regenerate, and selective application of suggestions—empowered users to experiment without fear of irreversible errors, fostering a sense of agency and engagement absent in baseline's workflows.
By selecting, reading, and experimenting with suggestions, users gain a more active and deliberate role in shaping the scene. 

\subsection{Comparison with Prior Work}
\systemname{} advances the field of AI-assisted spatial design by addressing a critical gap in existing systems: the lack of intermediate user intervention during scene generation. Unlike end-to-end approaches like HOLODECK or VRCopilot~\cite{zhang2024vrcopilot}, which limit user input to initial prompts or post-generation adjustments, \systemname{} externalizes the AI’s reasoning process through actionable suggestions. This aligns with emerging HCI paradigms that emphasize progressive co-creation outside the spatial design scenarios~\cite{10.1145/3491101.3519873, yuan2022wordcraft}, where users iteratively steer AI outputs rather than passively accepting results. Our findings echo prior work on AI-supported writing tools~\cite{yuan2022wordcraft}, where intermediate suggestions stimulate ideation, but extend these principles to 3D spatial design by integrating multimodal reasoning and immersive interaction.

\subsection{Design Implications}

\paragraph{\textbf{Intermediate Suggestions Enhance Creativity}} 
Exposing AI-generated suggestions as modular, interactable components helps users bridge the gap between abstract ideas and concrete implementations. This approach not only mitigates the ``blank canvas'' problem but also introduces serendipitous elements that spark new ideas.

\paragraph{\textbf{Flexible Control Mechanisms}} 
Features like undo and regenerate reduce the cognitive cost of experimentation of ideas, enabling users to explore divergent design paths without friction. Future systems should prioritize such reversible interactions to balance automation with user agency. More exploration of design approaches facilitating quasi-execution could also be promising. 

\paragraph{\textbf{Abstraction-Aware AI Pipelines}} \systemname{}’s strong performance on high-abstraction instructions suggests that generative AI models are more suitable for supporting tasks with relatively abstract goals. Vague prompts could trigger broader exploratory suggestions, while concrete requests might prioritize precision. While aligning mental models is hard for human-AI collaboration, involving users in the decision making process can facilitate idea buy-in and co-creation. Perhaps for tasks in medium abstraction levels, AI systems could shift the discussion with users between abstraction levels for intent alignment. 

\subsection{Limitations and Future Work}

While \systemname{} demonstrates promise, our studies highlighted areas for improvement. First is a potentially high cognitive load. Participants occasionally experienced decision fatigue when evaluating multiple suggestions. Future iterations could incorporate user intent modeling to prioritize or filter suggestions dynamically. Secondly, users desired finer control over which scene elements are modified by the AI (e.g., protecting manually adjusted objects). Implementing ``focus modes'' or exclusion zones could address this. In addition, participants also suggested integrating sketches or spatial gestures alongside voice commands to enrich expressiveness. These are all promising features to add in future systems. \rewrite{Moreover, there are other meaningful comparisons to evaluate our proposed system. For instance, comparing it with a similar system like LLM \cite{de2024llmr}, which shares conceptual similarities but differs in technical components, could help assess different implementation choices.}

\section{Conclusion}
This work designed, implemented and tested a novel system that enables progressive spatial design within the immersive VR environment. It differs from existing approaches by focusing on supporting iteration through AI-assisted modification rather than zero to one generation, which is achieved by enabling users to read and interact with the intermediate suggestions of AI automation. Our technical evaluation showed benefits of each of our pipeline component, while our user evaluation revealed benefits of providing this intermediate layer of interaction, including its better creativity support and user control. Our study also found that showing suggestions affected users' creative strategy by leaning more towards a top-down approach with global planning, while a baseline approach appeared more exploratory. Our findings underscore the value of progressive design workflows in immersive environments and provide a foundation for future systems that blend automation with embodied user agency.

% \textit{SceneCraft} has been shown to be a fast scene understanding and modification system which has got accolades from the users because of its imagination, still it has position misidentifications due to the small area of a shot it can analyze. It has a good ability to follow explicit instructions, but to some degree, it is fell short to process abstract commands, and thus quite random. The robot better interpreting the commands and broadening field of view could help increase the accuracy and consequently user satisfaction rate by a large margin.

%In this paper, we introduce \textit{SceneCraft}, an AI-assisted pipeline for virtual scenes modification by LVLMs. We focused on interior design scenarios, exploring the user command input space and assessing its effectiveness through two user studies. The results demonstrated that \textit{SceneCraft} can effectively interpret 3D scenes and various abstractions of user commands, generating suggestions and applying results that meet user expectations across functional, aesthetic, and psychological dimensions.

\begin{acks}
This project was funded by the National Natural Science Foundation of China - Young Scientists Fund (CityU 62202397), and the Key Project of the Institute of Software Chinese Academy of Sciences (ISCAS-ZD-202401). Special thanks to Ruishan Wu and Chenyue Guo for contributing to early explorations of this work and to Dr Teng Han for fruitful discussions. We also thank our reviewers for putting forward valuable suggestions and all the participants in our user studies.
\end{acks}

% \section{Experiment Design}
% \textcolor{blue}{Nianlong, Ruishan}

% User Instruction Collection

% \section{Evaluation Results}

% \begin{acks}
% \end{acks}

\bibliographystyle{ACM-Reference-Format}
\bibliography{sample-base}
\clearpage
\appendix
\section{Appendix}
%\zz{Add prompt and json file format in appendix}
\subsection{Details of Labeling Module}\label{Details of Labeling Module}
\quad\textbf{\textit{Prompt:}}

\begin{fbox}{
\begin{minipage}{0.45\textwidth}
\textbf{System Prompt}: Assume you're assisting users in automating picture labeling, You will receive a base64 code of a image. Based on all this data, generate the information of data as JSON format.

The format should like:

\{

    "name":"object\_name",

    "description":"object\_description",

    "category":"object\_category"

\}.

Here, I will offer you the object\_name, you should use it to generate the JSON. Description should only include the function, color, material, aesthetics and psychology of this object in the image, please use at most three simple sentences to finish the description, try to keep description very concise. Category is the category in reality of the object in the image. Categories such as "3D model", "3D shape" and so on are not be allowed. Do not generate extra string or information when you generate JSON.

\textbf{User Prompt}: object\_name: Armchair1\_C1

image: \textit{the base64 code of model image}.
\end{minipage}}
\end{fbox}

\textbf{\textit{JSON Format Object Annotation:}}

\begin{fbox}{
\begin{minipage}{0.45\textwidth}
\{

    "name": "Armchair1\_C1",
    
    "description": "This is a contemporary style armchair with a sleek black color finish, likely made of a material such as leather or synthetic upholstery. Its design is intended for comfortable seating with a modern aesthetic, potentially contributing to a sophisticated and minimalistic ambiance in a living space.",
    
    "category": "Chair"
    
\}
\end{minipage}}
\end{fbox}

\subsection{Details of Generative Module}\label{Details of Generative Module}

\quad\textbf{\textit{Scene Understanding:}}\label{Scene Understanding}

\begin{fbox}{
\begin{minipage}{0.45\textwidth}
\textbf{System Prompt}: I will give you a list of objects in json format, includes the names, coordinate points, rotation vectors, sizes of the objects, and hexadecimal color codes of objects in the 3D scene, also I will provide you the top view picture of the 3D scene, please understand this scene, please understand this scene.

\textbf{User Prompt}: Object list: \textit{JSON format objects' parameters}. 

Top View Image: \textit{the base64 code of top view image of scene.}    
\end{minipage}}
\end{fbox}

\textbf{\textit{Suggestions Generation:}}\label{Suggestions Generation}

\begin{fbox}{
\begin{minipage}{0.45\textwidth}
\textbf{System Prompt}: As a VR scene designer, you are presented with a detailed information of a 3D space scene. Your task is to interpret abstract user instructions for modifying this VR scene. Based on the scene's current layout, objects' attributes, and user commands, propose several creative and feasible suggestions for adjustments. These suggestions may involve repositioning furniture, altering object colors, adjusting sizes, or introducing 
\end{minipage}}
\end{fbox}

\begin{fbox}{
\begin{minipage}{0.45\textwidth}
new elements to enhance the space's functionality and aesthetic appeal. Ensure your proposals are clear, specific, and aligned with the user's desires, providing a blend of practicality and innovative design. Please provide modification suggestions and solutions with JSON format. For example, if you provide some suggestions, the result is:

\{

    "suggestions":[
    
        \{
        
            "suggestion":"add something and move something, change color"
            
        \},
        
        \{
            
            "suggestion":"add something and change color, also, change style"
            
        \},
        
        \{
            
            "suggestion":"change color, destroy something"
            
        \},
        
                                ....
                                
        \{
        
            "suggestion":"move something, rotate something"
            
                                \}]
                                
                            \}
                        
Each suggestions item can only include the suggestion, DO NOT include any other characters. Avoid extraneous text or characters outside the specified JSON format. The return format only includes JSON content, start with the first \{ of json.

\textbf{User Prompt}: 
User Instruction : \textit{User Instruction}

Object list: \textit{JSON format objects' parameters}. 

Top View Image: \textit{the base64 code of top view image of scene.}   
\end{minipage}}
\end{fbox}

\vspace{1em}
\begin{fbox}{
\begin{minipage}{0.45\textwidth}
\textbf{Suggestions JSON format}: 

\{

  "suggestions":[
  
    \{
    
      "suggestion":"add a large screen on Wall\_N for a cinema effect and install surround sound speakers around the room"
      
    \},
    
    \{
    
      "suggestion":"change the wall color to dark gray or black for an immersive cinema feel"
      
    \},
    
    \{
    
      "suggestion":"rearrange the room by adding comfortable recliner chairs in front of the screen"
   
    \},

....
    
    \{
    
      "suggestion":"adjust the ceiling height to accommodate a projector or large screen installation"
    
    \}]
  
\}
\end{minipage}}
\end{fbox}

\clearpage
\textbf{\textit{Actions Generation:}}\label{Actions Generation}

\begin{fbox}{
\begin{minipage}{0.45\textwidth}
\textbf{System Prompt:} Translate design suggestions into specific VR 3D space modifications based on JSON scene parameters. Output must strictly adhere to the JSON format below, detailing implementation steps for Add, Move, Rotate, Scale, Color, Style, and Destroy actions. You must remember DO NOT include other redundant text in the generated content, the return format only includes JSON content, start with the first "\{" of JSON:

\{

"steps": [\{
    
    "action": "Specify\_Action\_Name",
    
    "action\_command": "Action\_Name \{Object\_Name\} to [Modification\_Value]",
    
    "selected\_obj": "Object\_Name",
    
    "key": "Modification\_Value"
    
    \},
    
    ...

    \{
    
    "action": "Specify\_Action\_Name",
    
    "action\_command": "Action\_Name \{Object\_Name\} to [Modification\_Value]",
    
    "selected\_obj": "Object\_Name",
    
    "key": "Modification\_Value"
    
    \}]

\}

Notes: For Add Command: Set 'action\_name' to "Add", use the format "Add \{Object\} to [(Position)]", and provide "key" with Vector3 position in (0,0,0) format as "Modification\_Value". For Move Command: Use "Move \{Object\_Name\} to [(New\_Position)]" format. For Rotate Command: Use "Rotate \{Object\} [(Angle)]" format, specifying Vector3 angle in (0,0,0) format in "key". Make sure the back of objects facing the nearest wall. For Scale Command: Use "Scale \{TV\} [1.2] times", should specify scaling extent as an integer in "key". For Color Command: Use "Color \{Table\} to red[(255, 0, 0)]", color require RGB Vector3 in (0,0,0) format for Modification\_Value. For Style Command, Use "Change \{Table\} to [Wood]", "key" is the material type as a string, including Basket, Black\_Plastic, Brick, Bronze\_Metal, Copper\_metal, Dark\_Oak, Flow\_Water, Flower\_Pattern, Glass, Glass\_Dark, Golden\_metal\_material, Grass, Leaf\_Pattern, Leather, Marble, Rustic\_Wood, Shiny\_Metal. For Destroy Command: "Destroy \{Cup\}", need "selected\_obj", action command and key. If the object you want to manipulate does not exist in the scene, you will need to "Add" this object before you manipulate it. Do not forget \{\} and () Avoid extraneous text or characters outside the specified JSON format, the return format only includes json content, start with the first "\{" of JSON"

\textbf{User Prompt:} Suggestion : \textit{Suggestion}

Object list: \textit{JSON format objects' parameters}. 

Top View Image: \textit{the base64 code of top view image of scene.}
\end{minipage}}
\end{fbox}

\vspace{1em}
\begin{fbox}{
\begin{minipage}{0.45\textwidth}
\textbf{Actions JSON format:}

\{
  
  "steps": [
    
    \{
\end{minipage}}
\end{fbox}

\vspace{1em}
\begin{fbox}{
\begin{minipage}{0.45\textwidth}
      "action": "Add",
      
      "action\_command": "Add {Movie\_Poster} to [(-3.80, 1.00, 0.05)]",
      
      "selected\_obj": "Movie\_Poster",
      
      "key": "(-3.80, 1.00, 0.05)"
      
    \},
    
    ...
    
    \{
      
      "action": "Move",
      
      "action\_command": "Move {Movie\_Poster} to [(-1.00, 1.00, -3.95)]",
      
      "selected\_obj": "Movie\_Poster",
      
      "key": "(-1.00, 1.00, -3.95)"
      
    \}]

\}
\end{minipage}}
\end{fbox}

For "Add" action, \systemname{} sends LLM the object name and categories list from our 3D model asset. LLM selects appropriate category and description for the object to be added based on context. \systemname{} searches for the object that best matches the description generated by the LLM among the responding category and adds it to the scene. The specific prompt is as follows:

\begin{fbox}{
\begin{minipage}{0.45\textwidth}
\textbf{System Prompt: }

I will offer you a name of object, a list of categories, you should provide me with the perfect category that best fit the object and the description about the object, description should include the function, material, aesthetics and psychology of this object, please use at most three simple sentences to finish the description, try to keep description very concise.you give me categories you chosen and description as this JSON format:

\{

"Category1":"Category1",

"Description":"description"

\}

\textbf{User Prompt: }

The object is : object\_name.

Categories include: category\_list.

\end{minipage}}
\end{fbox}

\subsection{Statistical Data of Ablation Study}\label{Statistical Data of Ablation Study}

\begin{table}[hbtp]
\begin{tabular}{c|cccc}
\hline
\multirow{2}{*}{Category} & \multicolumn{4}{c}{Friedman Test}                    \\ \cline{2-5} 
                        & \textit{W} & \textit{df} & \textit{p}       & $\chi^2(3)$     \\ \hline
Relevance               & 0.315      & 3           & \textless{}0.001 & 148.53 \\ \hline
Preference              & 0.295      & 3           & \textless{}0.001 & 138.96 \\ \hline
Reasonableness           & 0.128      & 3           & \textless{}0.001 & 60.38  \\ \hline
Inspiration             & 0.242      & 3           & \textless{}0.001 & 113.91 \\ \hline
\end{tabular}
\caption{Friedman Test of scene modification with different components conditions.}
\label{FriedmanTestComponents}
\end{table}

\onecolumn
\begin{table}[hbtp]
\begin{tabular}{ccccc}
\hline
                                                 & Relevance                   & Preference                  & Reasonableness              & Inspiration                 \\ \hline
\textit{\systemname{} - V+OP+S} & $mean = 4.18$, $SD = 0.966$ & $mean = 3.86$, $SD = 1.100$ & $mean = 3.59$, $SD = 1.214$ & $mean = 3.80$, $SD = 1.185$ \\ \hline
\textit{OP+S}                                    & $mean = 3.37$, $SD = 1.402$ & $mean = 3.01$, $SD = 1.394$ & $mean = 2.99$, $SD = 1.441$ & $mean = 2.99$, $SD = 1.441$ \\ \hline
\textit{V+OP}                                    & $mean = 3.24$, $SD = 1.237$ & $mean = 2.89$, $SD = 1.240$ & $mean = 2.76$, $SD = 1.361$ & $mean = 2.76$, $SD = 1.361$ \\ \hline
\textit{V+S}                                     & $mean = 2.40$, $SD = 1.192$ & $mean = 2.21$, $SD = 1.155$ & $mean = 2.39$, $SD = 1.433$ & $mean = 2.35$, $SD = 1.187$ \\ \hline
\end{tabular}
\caption{The mean score and $SD$ of each \emph{Input Configuration} in different categories.}
\label{InputMeanTable}
\end{table}

\begin{table}[hbtp]
\begin{tabular}{ccccc}
\hline
                & Relevance                  & Preference                 & Reasonableness             & Inspiration                \\ \hline
\textit{Low}    & $mean = 4.32$, $SD = 1.08$ & $mean = 3.94$, $SD = 1.17$ & $mean = 3.92$, $SD = 1.23$ & $mean = 3.40$, $SD = 1.36$ \\ \hline
\textit{Medium} & $mean = 3.86$, $SD = 0.97$ & $mean = 3.52$, $SD = 1.16$ & $mean = 3.12$, $SD = 1.08$ & $mean = 3.68$, $SD = 1.11$ \\ \hline
\textit{High}   & $mean = 4.34$, $SD = 0.85$ & $mean = 4.12$, $SD = 0.96$ & $mean = 3.60$, $SD = 1.23$ & $mean = 4.32$, $SD = 0.82$ \\ \hline
\end{tabular}
\caption{The mean score and $SD$ of different abstraction levels in different categories, in this table Low, Medium, and High are Low Abstraction, Medium Abstraction and High Abstraction.}
\label{AbstractionLevleMean}
\end{table}

\begin{table}[hbtp]
\centering
\resizebox{0.46\textwidth}{!}{  % 开始缩放整个表格
\begin{tabular}{c|c|cccc}
\hline
\multirow{2}{*}{Category} & \multirow{2}{*}{Abstraction Level} & \multicolumn{4}{c}{Friedman Test} \\ \cline{3-6} 
                          &                                    & \textit{W} & \textit{df} & \textit{p} & $\chi^2(2)$ \\ \hline
Relevance                 & L×M×H                              & 0.099      & 2           & 0.007      & 9.94        \\ \hline
Preference                & L×M×H                              & 0.094      & 2           & 0.009      & 9.373       \\ \hline
Reasonableness            & L×M×H                              & 0.109      & 2           & 0.004      & 10.88       \\ \hline
Inspiration               & L×M×H                              & 0.165      & 2           & $<$0.001   & 16.513      \\ \hline
\multirow{2}{*}{Category} & \multirow{2}{*}{Abstraction Level} & \multicolumn{4}{c}{Wilcoxon signed-rank tests} \\ \cline{3-6} 
                          &                                    & \textit{W} & \textit{Z}  & \textit{p} & \textit{r}  \\ \hline
\multirow{3}{*}{Relevance}     & L×M & 180 & 2.800 & 0.015 & 0.396 \\ \cline{2-6}
                               & L×H & 175 & -0.026 & 1.0 & 0.004 \\ \cline{2-6}
                               & M×H & 145 & -2.568 & 0.031 & 0.363 \\ \hline
\multirow{3}{*}{Preference}    & L×M & 217 & 2.103 & 0.106 & 0.297 \\ \cline{2-6}
                               & L×H & 241 & -0.724 & 0.468 & 0.102 \\ \cline{2-6}
                               & M×H & 165 & -2.702 & 0.021 & 0.382 \\ \hline
\multirow{3}{*}{Reasonableness} & L×M & 165 & 3.362 & 0.002 & 0.475 \\ \cline{2-6}
                                & L×H & 297 & 1.327 & 0.554 & 0.188 \\ \cline{2-6}
                                & M×H & 265 & 1.994 & 0.046 & 0.282 \\ \hline
\multirow{3}{*}{Inspiration}   & L×M & 319 & -1.100 & 0.814 & 0.177 \\ \cline{2-6}
                               & L×H & 120 & -3.550 & $<$0.001 & 0.545 \\ \cline{2-6}
                               & M×H & 107 & 2.450 & 0.043 & 0.453 \\ \hline
\end{tabular}
}  % <- 结束 resizebox
\caption{Statistical data of scene modification with different abstraction levels. In this table, L, M, H are Low, Medium, and High Abstraction.}
\label{AbstractionLevelTable}
\end{table}

\begin{figure}[htbp]
\centering
 \includegraphics[width=0.5\textwidth]{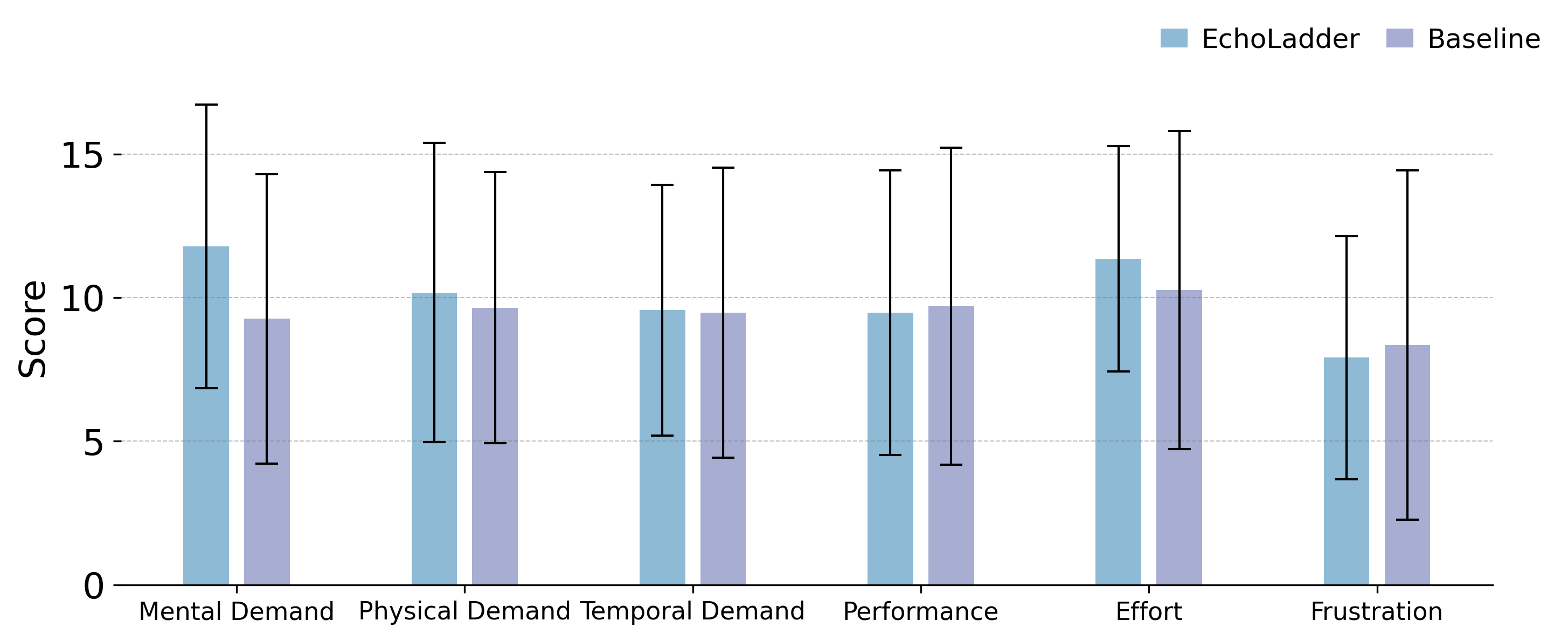}
\caption{NASA TLX results for \systemname{} and baseline (Full score is 21).}
  \label{NASATLX}
\end{figure}

\twocolumn

\clearpage

\subsection{Participants' Workflow}
\begin{figure}[htbp]
\centering
	\subfigure{\includegraphics[width=0.8\textwidth]{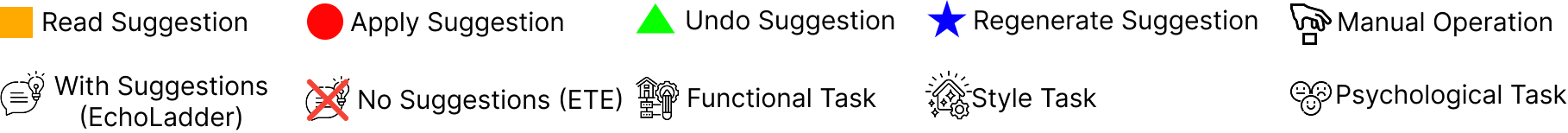}}
        \subfigure{\includegraphics[width=1\textwidth]{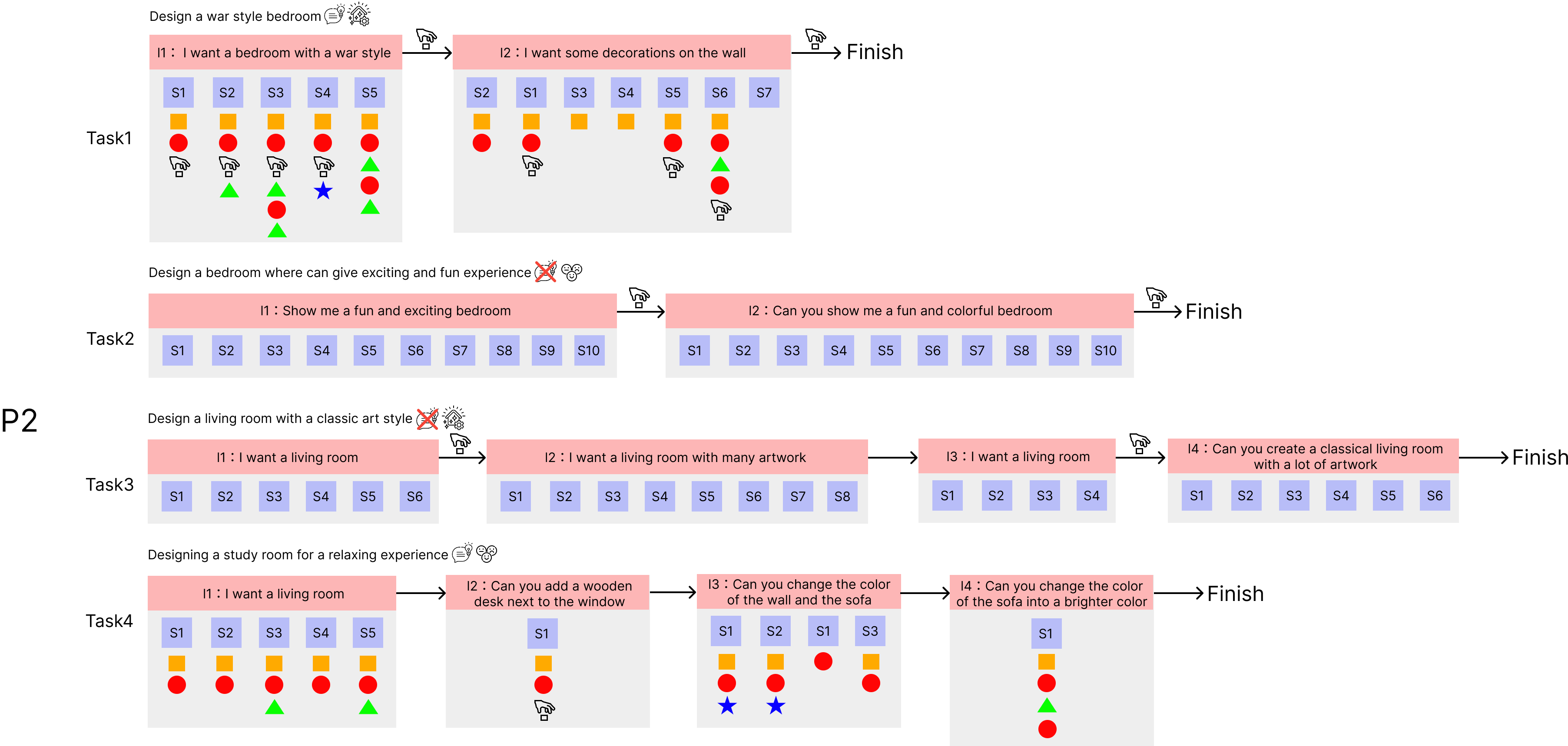}}
\end{figure}

\begin{figure*}
\centering
        \subfigure{\includegraphics[width=1\textwidth]{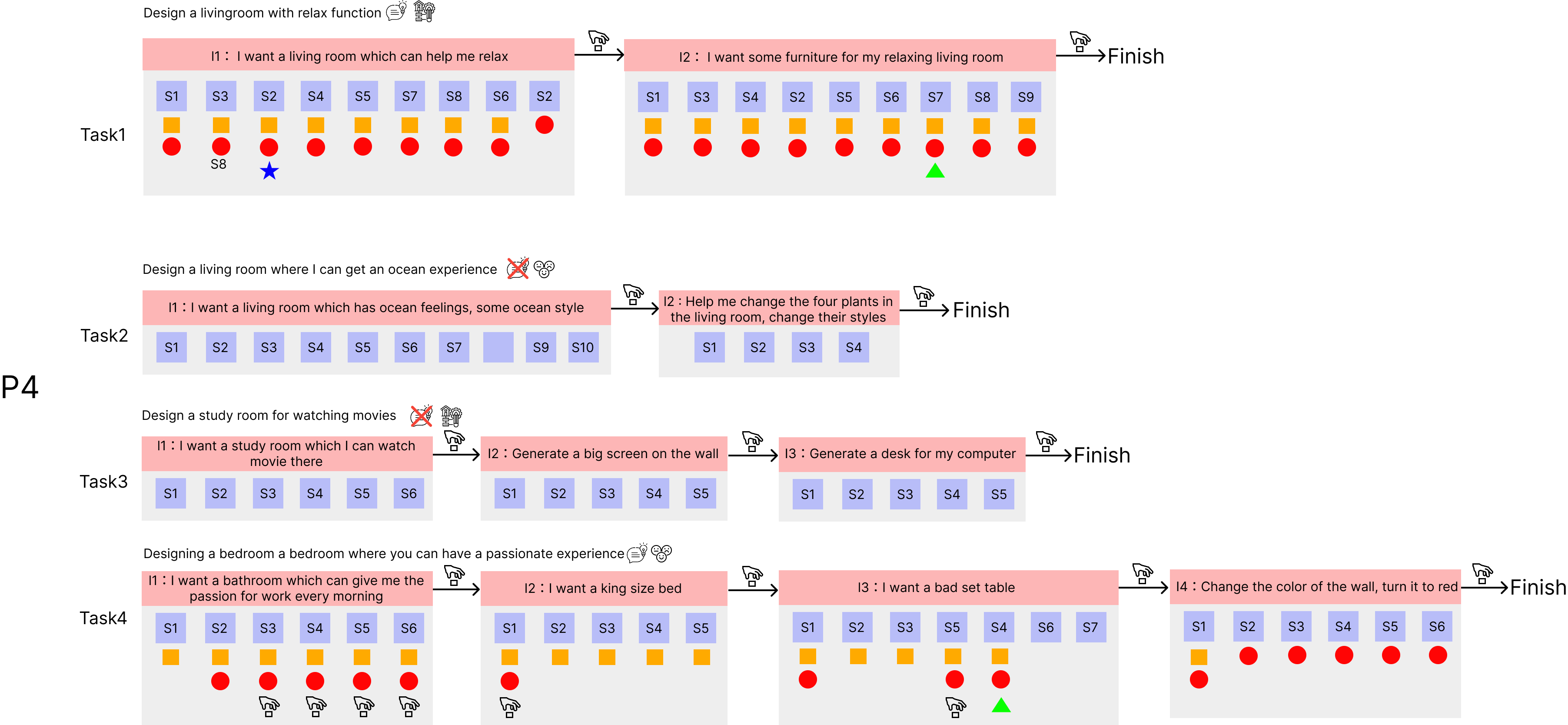}}
\end{figure*}

\begin{figure*}[htbp]
\centering
        \subfigure{\includegraphics[width=1\textwidth]{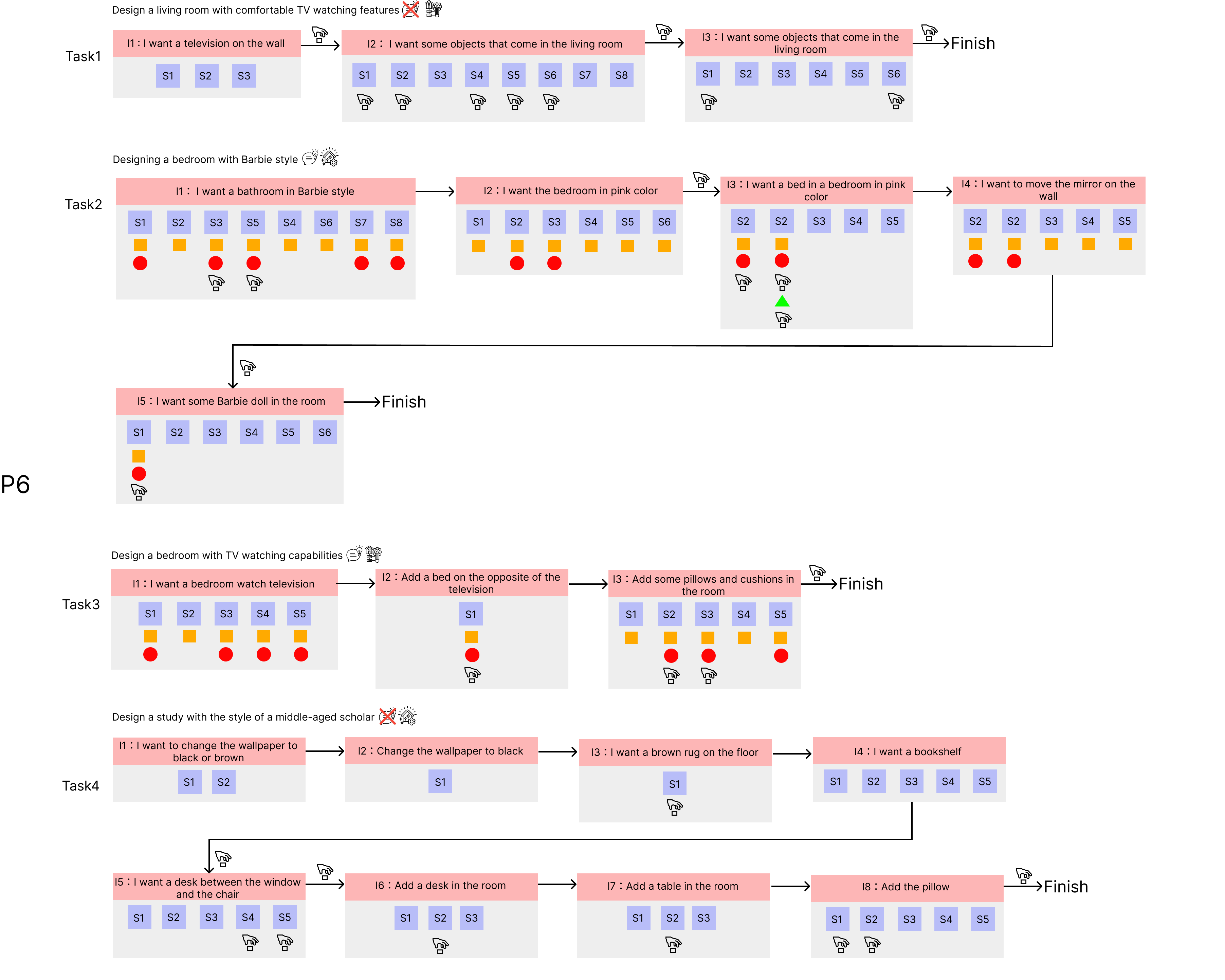}}
\end{figure*}

\begin{figure*}[htbp]
\centering
        \subfigure{\includegraphics[width=1\textwidth]{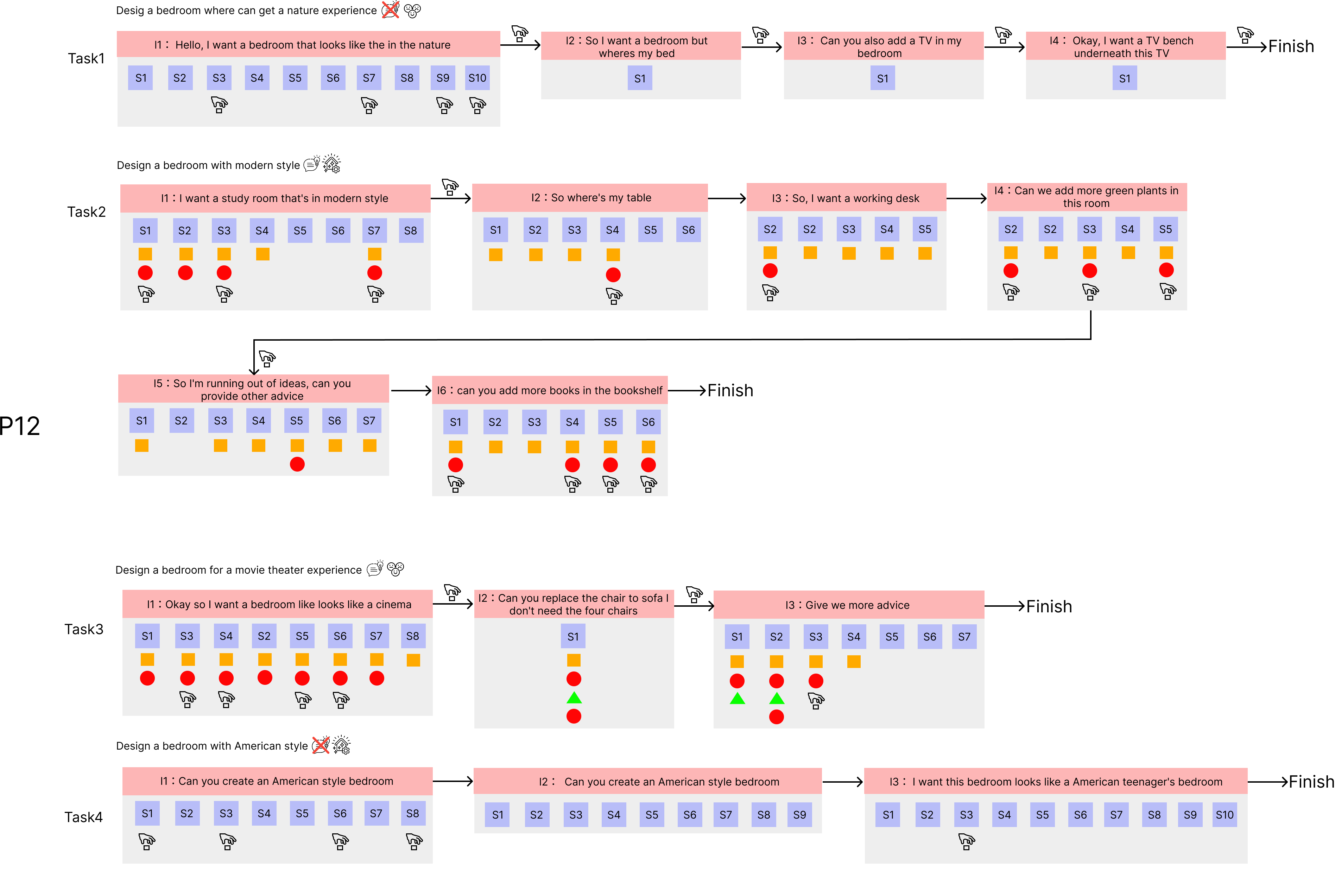}}
\caption{Example participants' workflows (P2, P4, P6, P12).}
  \label{Workflow}
\end{figure*}

\end{document}